\documentclass[preprint,12pt]{elsarticle}

\usepackage{epsfig}
\usepackage{enumerate}

\usepackage{amssymb,amsmath,graphics,graphicx}  
\usepackage[dvipsnames]{xcolor}

\journal{Elsevier}

\begin{document}

\begin{frontmatter}


\title{Nonequilibrium phase transitions and absorbing states in a  model for the dynamics of religious affiliation}

\author{Nuno Crokidakis}
\ead{nunocrokidakis@id.uff.br}

\address{Instituto de F\'{\i}sica, Universidade Federal Fluminense, Niter\'oi, Rio de Janeiro, Brazil}

\begin{abstract}
  We propose a simple model to describe the dynamics of religious affiliation. For such purpose, we built a compartmental model with three distinct subpopulations, namely religious committed individuals, religious noncommitted individuals and not religious affiliated individuals. The transitions among the compartments are governed by probabilities, modeling social interactions among the groups and also spontaneous transitions among the compartments. First of all, we consider the model on a fully-connected network. Thus, we write a set of ordinary differential equations to study the evolution of the subpopulations. Our analytical and numerical results show that there is an absorbing state in the model where only one of the subpopulations survive in the long-time limit. There are also regions of parameters where some of the subpopulations coexist (two or three). We also verified the occurrence of two distinct critical points. In addition, we also present Monte Carlo simulations of the model on two-dimensional square lattices, in order to analyze the impact of the presence of a lattice structure on the critical behavior of the model. Comparison of the models' results with data for religious affiliation in Northern Ireland shows a good qualitative agreement. Finally, we considered the presence of inflexible individuals in the population, i.e., individuals that never change their states. The impact of such special agents on the critical behavior of the model is also discussed.

\end{abstract}


\end{frontmatter}



\section{Introduction}

The decline of religious affiliation has been observed in some countries recently. For example, in USA the decline in church membership is primarily a function of the increasing number of Americans who express no religious preference. A 2021 poll shown that over the past two decades, the percentage of Americans who do not identify with any religion has grown from $8\%$ in 1998-2000 to $13\%$ in 2008-2010 and $21\%$ in 2019-2021 \cite{gallup}. In Europe, it has been observed an accelerated decline in religious belief since 1980 \cite{dogan}.

The study of the dynamics of religion evolution was attracted the attention of researchers through the world \cite{ausloos1,ausloos2,picoli,abrams,glass,labzai}. Considering religions as population dynamics, the authors in \cite{ausloos1} considered Weibull and lognormal functions to fit the distribution of number of adherents of religions. The same authors also considered a statistical physics like approach to the problem \cite{ausloos2}. Another study considered the growth dynamics of religious activities. It was found that the distribution of annual logarithmic growth rates exhibits a non-Gaussian behavior, being well described by a Laplace distribution, for distinct size scales, and  that the standard deviation of the growth rates scales with size as a power law \cite{picoli}. In order to explain the decline of religious affiliation in some countries, some works considered models based on coupled differential equations \cite{abrams,glass}. In opposition to such mentioned studies, a recent work considered a discrete modeling of the propagation of religious ideas in order to maximize the number of religious individuals and minimize the individuals who renounce religion and do not practise religious rites, considering an optimal control strategy \cite{labzai}. 

Compartmental models are simple and important tools concerning the modeling of a diverse set of social dynamics, including the addoption of social habits \cite{glass,meu1,meu2,meu3,meu4,meu5,serge_javarone,fantaye,mamo,rahman,sooknanan,tax_evasion}. Following the ideas of references \cite{abrams,glass}, we consider in this work a three-state compartmental model to study the dynamics of religious affiliation. The model takes into account subpopulations $X$ (religious committed individuals), $Y$ (not religious affiliated individuals) and $Z$ (religious noncommitted individuals). Our target is to study the critical behavior of the model. Considering the model defined on fully-connected networks and on 2d square lattices, we found that the system presents two distinct critical points. These critical points separates three distinct regions, with distinct long-time behaviors. One of the regions represents an absorbing state where the dynamics becomes frozen. After the analysis of such model, we consider the presence of inflexible/zealot individuals in the subpopulations of religious committed individuals ($\textbf{X}$) and not religious ones ($\textbf{Y}$). Inflexible, or zealot, individuals are the ones that never change their status - in this case, either $\textbf{X}$ or $\textbf{Y}$ individuals will remain in their states during all the dynamics. This realistic situation can model the presence of radical individuals concerning religions, and the presence of inflexible individuals can dramastically change the dynamics and the critical behavior of social models \cite{drink,galam_inflex,nuno_jorge,galam1991,martins,celia_vitor,mobilia1,latoski,galla,verma,donner,mobilia2}.

This work is organized as follows. In section 2 we present the initial formulation of the model, in the absence of inflexible individuals in the population. We divide section 3 in two subsections, one to discuss analytical and numerical results for such initial model defined on a fully-connected network, and other to discuss results of Monte Carlo simulations of the model defined on the top of two-dimensional square lattices. In section 4, we consider the presence of zealot individuals in the population, and study the impact of such presence in the critical behavior of the initial model in section 5. Finally, in section 6 we present our final remarks.


\section{Model I: Three-state model with no inflexible agents}

Based on the works \cite{abrams,glass}, we consider a population of $N$ individuals or agents, where each agent $i$ ($i=1, 2, ..., N$) can be in one of three possible states, namely the subpopulations $X$ (religious committed individuals), $Y$ (not religious affiliated individuals) and $Z$ (religious noncommitted individuals). It can be argued that the consideration of such religious noncommitted group ($Z$) instead of only two groups ($X$ and $Y$) is a realistic feature for built models to study dynamics of religious affiliations. Indeed, the majority of Europe's Christians are non-practicing, but they differ from religiously unaffiliated people in their views on God, attitudes toward Muslims and immigrants, and opinions about religion’s role in society \cite{pew_research}.

Since we are considering a social contagion model, the probabilities related to changes in agents' compartments represent the possible contagions. The transitions among compartments are as following:

\begin{itemize}
\item $X \stackrel{p}{\rightarrow} Z$: a religious committed agent (\textbf{X}) becomes a religious noncommitted individual (\textbf{Z}) with probability $p$;

\item $Z \stackrel{q}{\rightarrow} X$: a religious noncommitted agent (\textbf{Z}) becomes a religious committed individual (\textbf{X}) with probability $q$ if he/she is in contact with religious committed individuals (\textbf{X});

\item $Z \stackrel{u}{\rightarrow} Y$: a religious noncommitted agent (\textbf{Z}) becomes a not religious affiliated individual (\textbf{Y}) with probability $u$ if he/she is in contact with not religious affiliated individuals (\textbf{Y});

\item $Y \stackrel{w}{\rightarrow} X$: a not religious affiliated agent (\textbf{Y}) becomes a religious committed individual (\textbf{X}) with probability $w$ if he/she is in contact with religious committed individuals (\textbf{X});

\end{itemize}

In the above rules, $p$ represents a spontaneous transitions, for which religious committed agents can lose interest in in attending churches or temples, but if they are questioned, they affirm they belong to a given religion (any religion). Thus, religious committed agents (\textbf{X}) can become religious noncommitted individuals (\textbf{Z}). The probability $q$ represents an ``infection'' probability, i.e., the probability that a religious noncommitted agent (\textbf{Z}) becomes a religious committed individual (\textbf{X}). We argue that the $\textbf{Z} \to \textbf{X}$ transition is more difficult to occur in comparison with $\textbf{X} \to \textbf{Z}$ transition, and for such reason we considered that the $\textbf{Z} \to \textbf{X}$ transition is due to to social pressure of religious committed agents (\textbf{X}) over religious noncommitted individuals (\textbf{Z}). In addition,  religious noncommitted individuals (\textbf{Z}) can be influenced by  not religious affiliated individuals (\textbf{Y}) to become  not religious affiliated individuals (\textbf{Y}). This transition is ruled by a social pressure probability $u$. Finally,  not religious affiliated individuals (\textbf{Y}) can become religious committed individuals (\textbf{X}) due to social pressure of religious committed individuals (\textbf{X}). This pressure is modeled by a probability $w$.

In the next section we will discuss our results.


\section{Results: Model I}

\subsection{Fully-connected population}

Considering the above-mentioned rules and a population defined on a fully-connected network, we can write the following mean-field equations for the compartments' evolution:
\begin{eqnarray} \label{eq1}
\frac{dx(t)}{dt} & = & -p\,x(t) + q\,x(t)\,z(t) + w\,x(t)\,y(t) \\  \label{eq2}
\frac{dy(t)}{dt} & = & -w\,x(t)\,y(t) + u\,y(t)\,z(t) \\ \label{eq3}
\frac{dz(t)}{dt} & = & -u\,y(t)\,z(t) - q\,x(t)\,z(t) + p\,x(t) 
\end{eqnarray}

\noindent

For simplicity, we consider a fixed population, i.e., at each time step $t$ we have the normalization condition $x(t)+y(t)+z(t)=1$, where we defined the population densities $x(t)=X(t)/N$, $y(t)=Y(t)/N$ and $z(t)=Z(t)/N$.

\begin{figure}[t]
\begin{center}
\vspace{6mm}
\includegraphics[width=0.45\textwidth,angle=0]{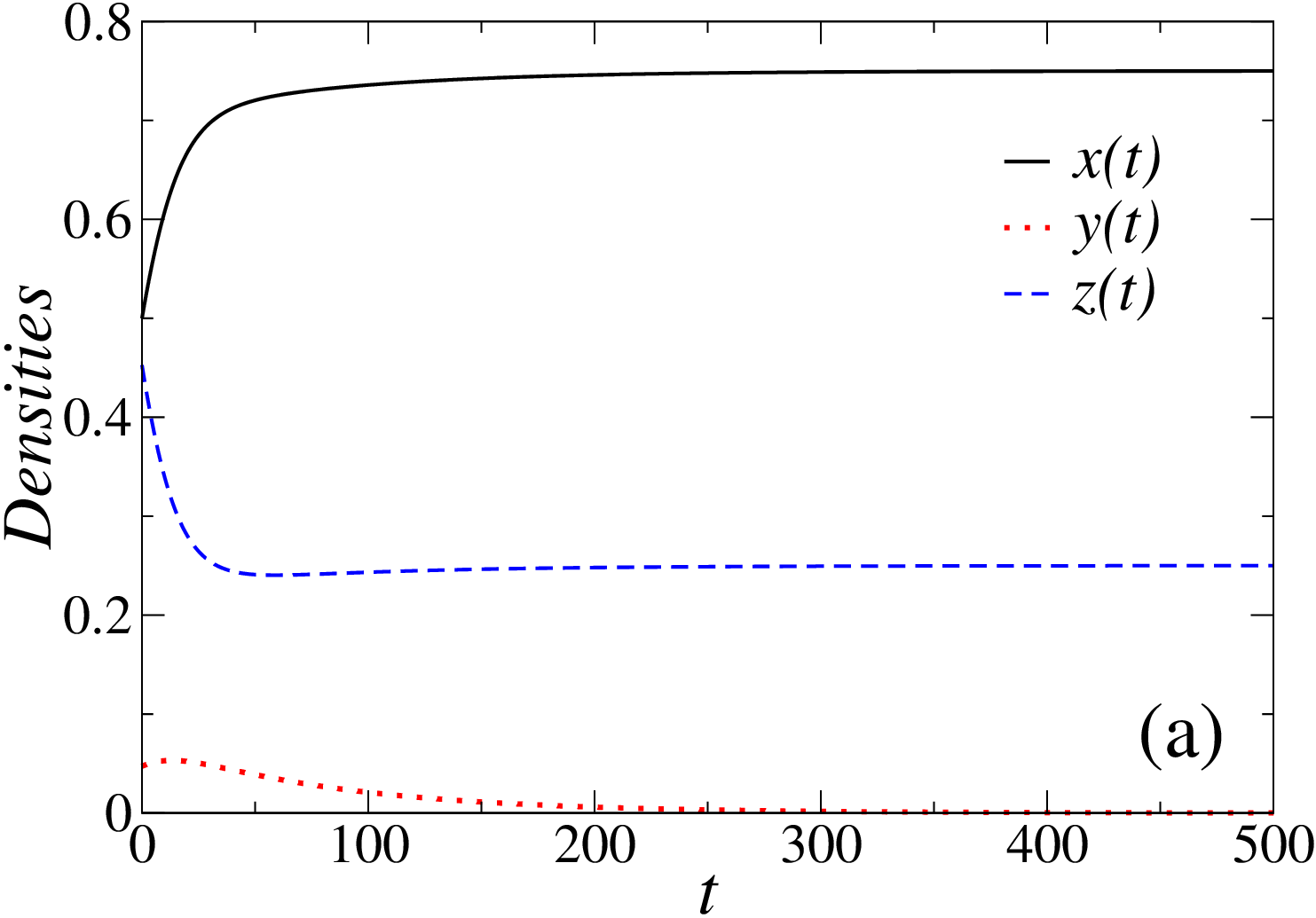}
\hspace{0.3cm}
\includegraphics[width=0.45\textwidth,angle=0]{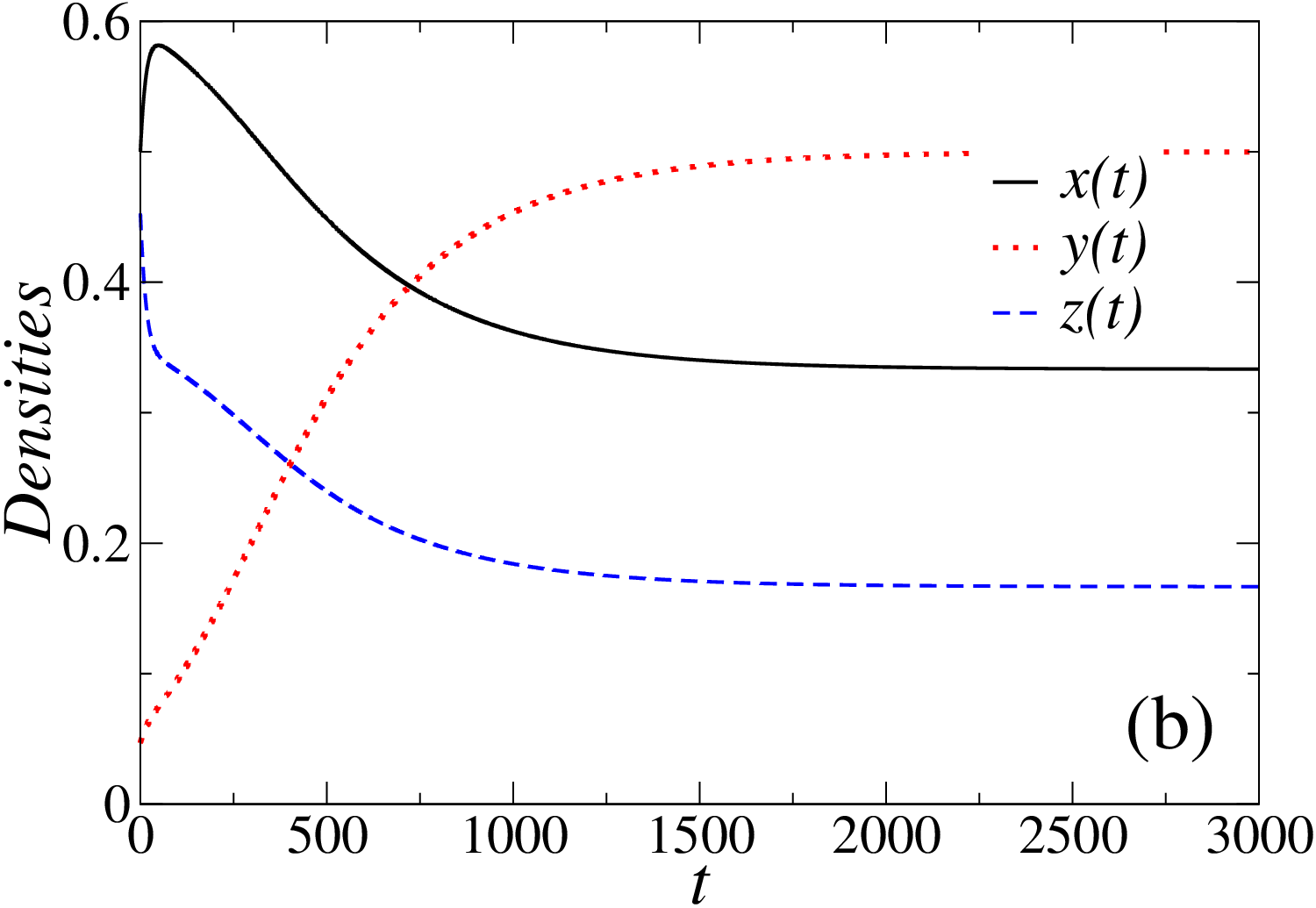}
\\
\vspace{1.0cm}
\includegraphics[width=0.45\textwidth,angle=0]{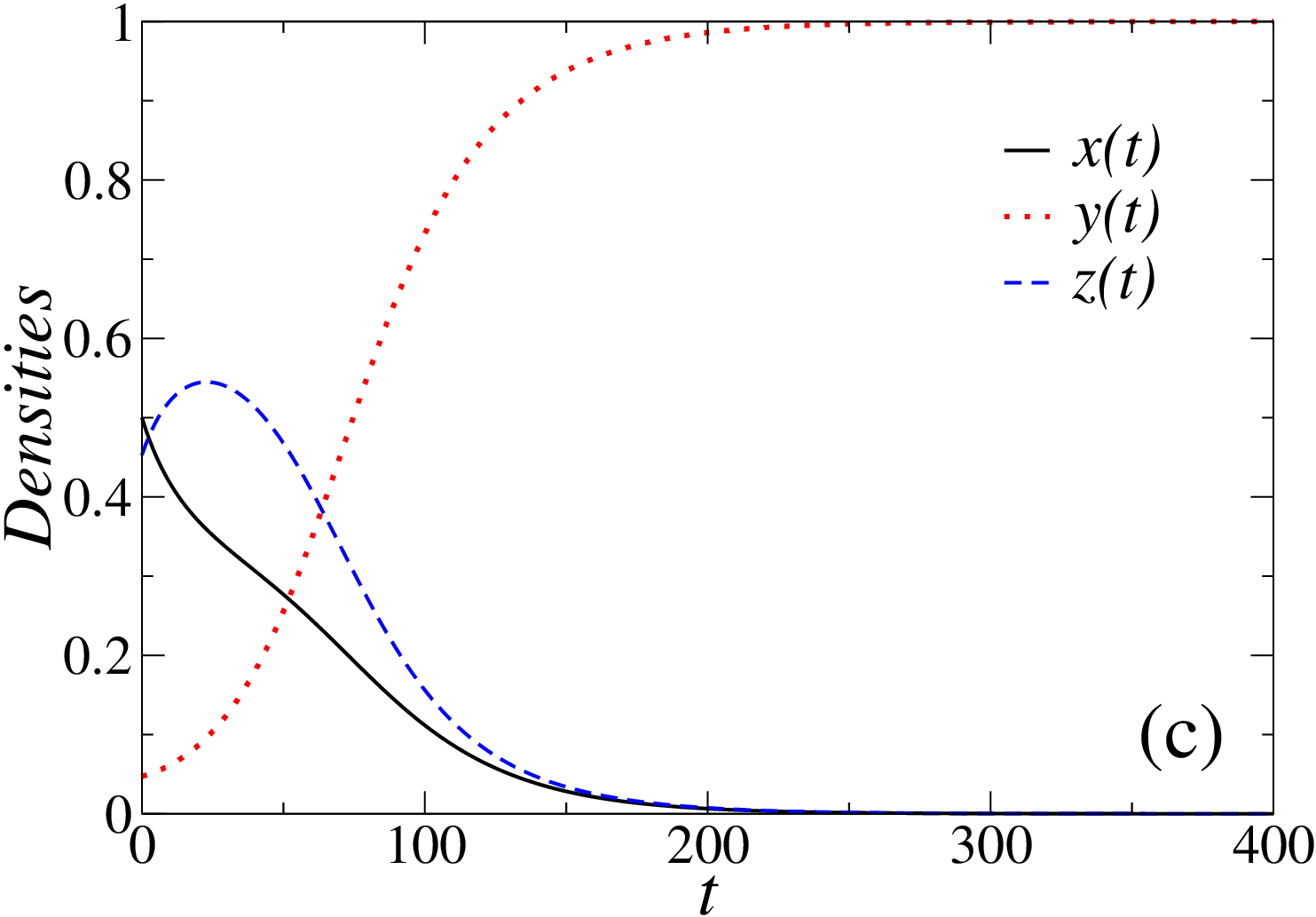}
\hspace{0.3cm}
\includegraphics[width=0.45\textwidth,angle=0]{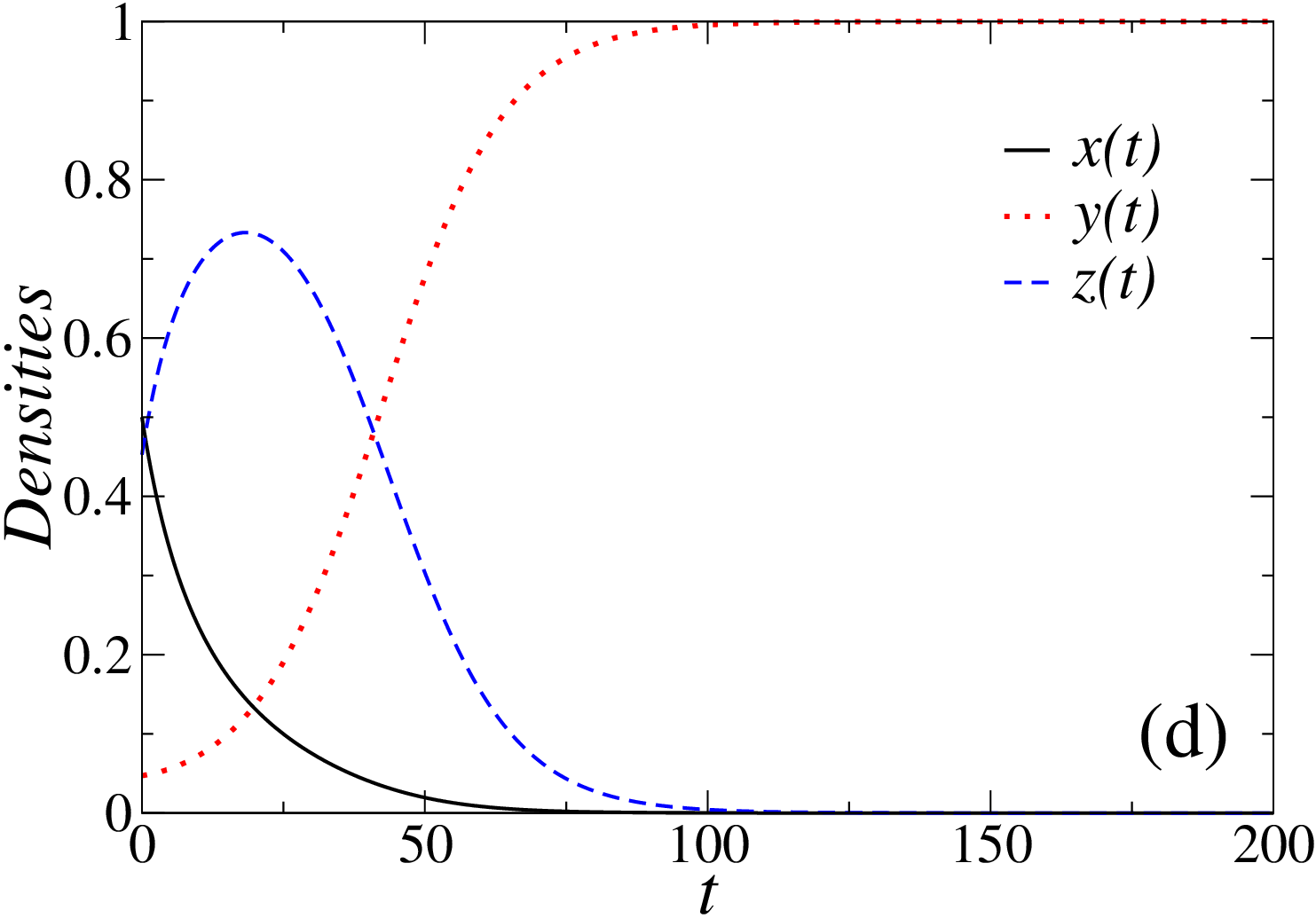}
\end{center}
\caption{(Color online) Time evolution of the three densities of agents $x(t)$, $y(t)$ and $z(t)$ for the mean-field formulation of the model, based on the numerical integration of Eqs. \eqref{eq1} - \eqref{eq3}. The fixed parameters are $w=0.05$, $q=0.12$ and $u=0.10$, and we varied the parameter $p$: (a) $p=0.030$, (b) $p=0.045$, (c) $p=0.080$, (d) $p=0.150$.}
\label{fig1}
\end{figure}

Let us start analyzing the time evolution of the previous defined quantities $x, y$ and $z$. We numerically integrated the Eqs. \eqref{eq1} - \eqref{eq3}, and the results are exhibited in Figure \ref{fig1}. For simplicity we fixed $w=0.05$, $q=0.12$ and $u=0.10$, and vary the parameter $p$. One can see that the increase of $p$ causes the increase of $y$ and the decrease of $x$ and $z$. It is also interesting to observe that, for sufficient high values of $p$ like $p=0.08$ (panel (c)) and $p=0.15$ (panel (d)), the densities evolve in time, and in the steady states we observe the survival of only the not religious affiliated subpopulation $y$, i.e., for $t\to\infty$ we have $x=z=0$ and $y=1$. This last result will be discussed in more details analytically in the following.

As we observed in Figure \ref{fig1}, the quantities $x(t), y(t)$ and $z(t)$ evolve in time, and after some time they stabilize. In such stationary states, the time derivatives of Eqs. \eqref{eq1} - \eqref{eq3} are zero. Denoting the stationary densities as $x=x(t\to\infty)$, $y=y(t\to\infty)$ and $z=z(t\to\infty)$, one can obtain three distinct solutions for the stationary states, namely (see Appendix),
\begin{align} \label{eq4}
x  =
\begin{cases} 1 - \frac{p}{q} & if \ p < p_{c}^{(1)}   
\\  
\frac{u\,(w-p)}{w\,(u+w-q)} &
if \ p_{c}^{(1)}<p<p_{c}^{(2)}
\\ 
0 & if \ p > p_{c}^{(2)} 
\end{cases}
\end{align}

\begin{align} \label{eq5}
y  =
\begin{cases} 0 & if \ p < p_{c}^{(1)}   
\\  
1-\frac{(u+w)\,(w-p)}{w\,(u+w-q)} & 
if \ p_{c}^{(1)}<p<p_{c}^{(2)}
\\ 
1 & if \ p > p_{c}^{(2)} 
\end{cases}
\end{align}

\begin{align} \label{eq6}
z  =
\begin{cases} \frac{p}{q} & if \ p < p_{c}^{(1)}   
\\  
\frac{w-p}{u+w-q} &
if \ p_{c}^{(1)}<p<p_{c}^{(2)}
\\ 
0 & if \ p > p_{c}^{(2)} 
\end{cases}
\end{align}
where the critical points are given by \footnote{The stability analysis of the three solutions \eqref{eq4} - \eqref{eq6} is discussed in the Appendix.}
\begin{eqnarray} \label{eq7}
p_{c}^{(1)} & = & \frac{w\,q}{u+w} \\ \label{eq8}
p_{c}^{(2)} & = & w   
\end{eqnarray}

Taking $y$ as the order parameter of the model (see the Appendix), the solution for $p>p_{c}^{(2)}$, i.e., $(x,y,z)=(0,1,0)$, defines an absorbing state where there are only not religious affiliated individuals in the population for long times. Looking for Fig. \ref{fig1} again, this absorbing state can be seen in panels (c) and (d) for long times. For the values of the considered parameters, namely $w=0.05$, $q=0.12$ and $u=0.10$, we obtain from Eqs. \eqref{eq7} and \eqref{eq8} the values $p_{c}^{(1)}=0.04$ and $p_{c}^{(2)}=0.05$. Thus for the considered values ($p=0.08$ in panel (c) and $p=0.15$ in panel (d)), the system evolves to steady states where $x=z=0$ and $y=1$, as predicted analytically in Eqs. \eqref{eq4} - \eqref{eq6}.

\begin{figure}[t]
\begin{center}
\vspace{3mm}
\includegraphics[width=0.7\textwidth,angle=0]{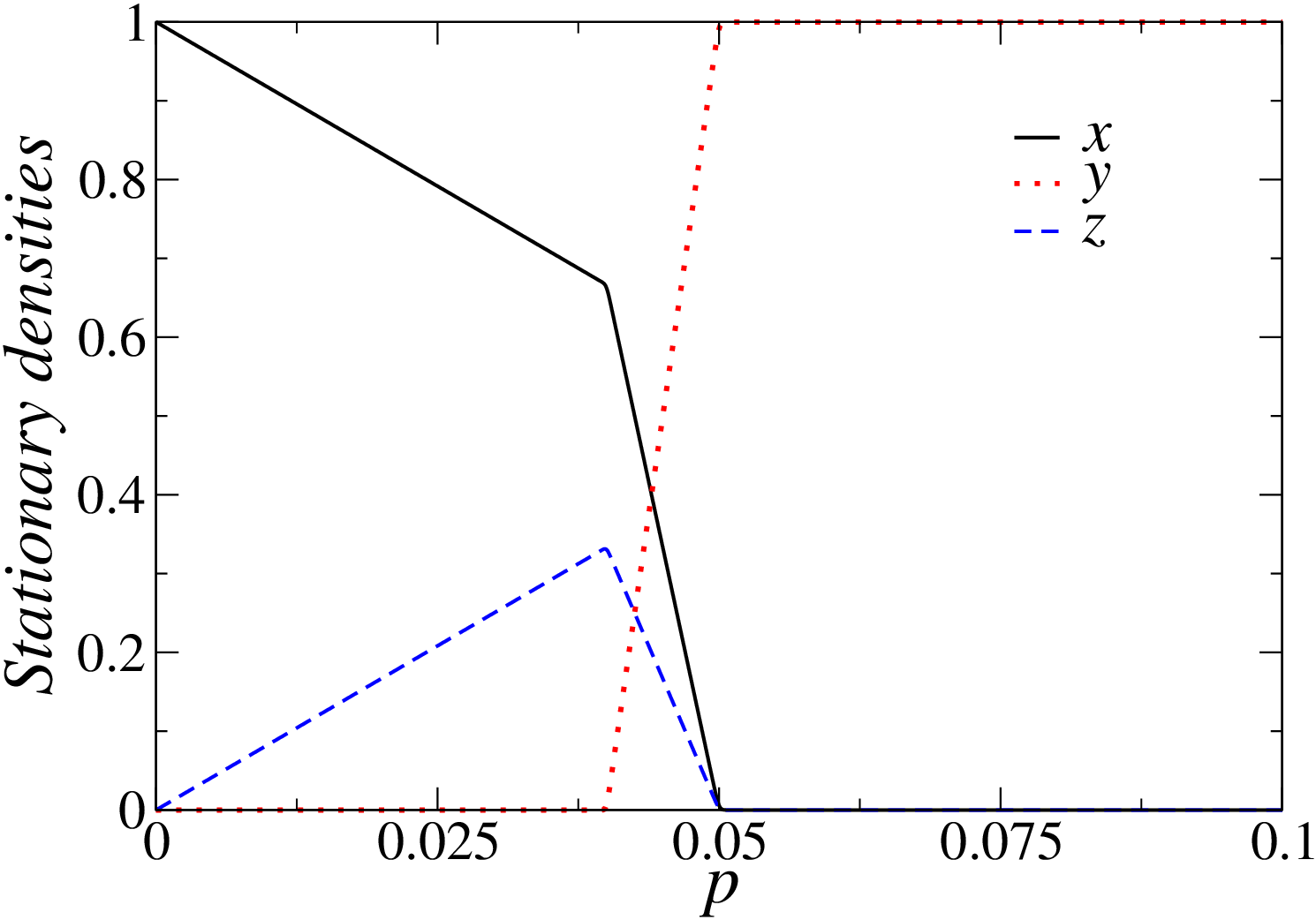}
\end{center}
\caption{(Color online) Stationary densities $x$, $y$ and $z$ as functions of the probability $p$ for $w=0.05$, $q=0.12$ and $u=0.10$ for the mean-field formulation of the model. The lines were obtained from Eqs. \eqref{eq4} - \eqref{eq6}. For the considered parameters, the critical points are $p_{c}^{(1)}=0.04$ and $p_{c}^{(2)}=0.05$. The absorbing state where $x=z=0$ and $y=1$ is observed for $p>p_{c}^{(2)}=0.05$. It is important to note that these behaviors are present also for other parameter values, and what is shown here works as a pattern.}
\label{fig2}
\end{figure}

Based on the results above, we plot in Fig. \ref{fig2} the steady state values of the three subpopulation densities $x$, $y$ and $z$ as functions of $p$. For this graphic, again we fixed the parameters $w=0.05$, $q=0.12$ and $u=0.10$. For such values, we have $p_{c}^{(1)}=0.04$ and $p_{c}^{(2)}=0.05$. As discussed above, for $p>p_{c}^{(2)}$ the system becomes frozen in the absorbing phase, where there are only not religious affiliated agents $y$ in the population after a long time, i.e., $y=1$ and $x=z=0$. For $p<p_{c}^{(1)}$ we have a region where the populations of religious individuals coexist, i.e., we have the religious committed and religious noncommitted individuals coexisting in the stationary states, whereas the population of not religious agents disappears. The realistic region where the three populations $x, y$ and $z$ coexist in the steady states is obtained for $p_{c}^{(1)} < p < p_{c}^{(2)}$.

\begin{figure}[t]
\begin{center}
\vspace{3mm}
\includegraphics[width=0.7\textwidth,angle=0]{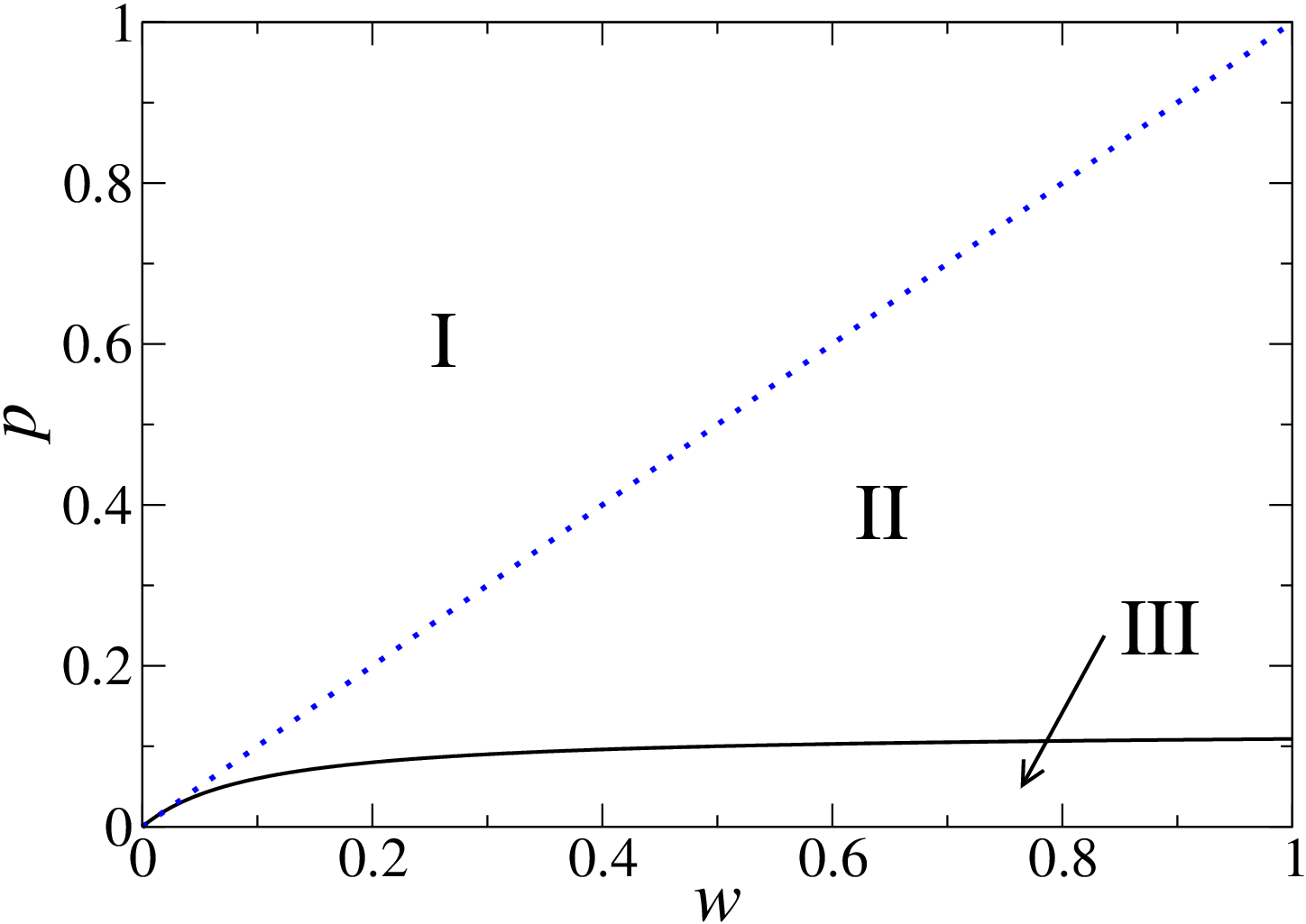}
\end{center}
\caption{(Color online) Phase diagram of the mean-field model in the plane $p$ \textit{vs} $w$ for $q=0.12$ and $u=0.10$. The full line represents the critical point $p_c^{(1)}$ given by Eq. \eqref{eq7} and the dotted line the other critical point $p_{c}^{(2)}$ given by Eq. \eqref{eq8}. \textbf{I} denotes the region where the the system falls in the absorbing phase $y=1, x=z=0$, \textbf{II} denotes the region where the three densities $x$, $y$ and $z$ coexist and \textbf{III} denotes the region where the subpopulation $y$ disappears and we observe the coexistence of subpopulations $x$ and $z$.}
\label{fig3}
\end{figure}

The competition among the contagions cause the occurrence of such three regions in the model \cite{drink}. From one side we have religious committed individuals influencing religious noncommitted ones to increase their beliefs about the religion, with probability $q$. On the other hand, we have the social pressure of not religious affiliated individuals over religious noncommitted ones, with probability $u$, in order to make such religious noncommitted agents to became not religious affiliated. We also have another social pressure of the religious committed individuals over not religious affiliated ones, with probability $w$, that intend to convert such not religiuos affiliated agents into religious commited ones. Finally, it is important to mention the parameter $p$, that drives the only transition of the model that does not depend on a direct social interaction. That parameter models the spontaneous decrease of interest in the religion of the individuals. Notice also that the parameter $w$ is present in both expressions for the nonequilibrium critical points $p_{c}^{(1)}$ and $p_{c}^{(2)}$, suggesting the importance of such parameter for the dynamics of the model.

For clarity, we exhibit in Fig. \ref{fig3} the phase diagram of the model in the plane $p$ versus $w$, separating the three above discussed regions. In Fig. \ref{fig3}, the absorbing phase with $y=1$ and $x=z=0$ is located in region I for $p<p_{c}^{(1)}$, the coexistence phase is denoted by II for $p_{c}^{(1)}<p<p_{c}^{(2)}$ (where the three densities $x, y$ and $z$ coexist) and region III denotes the region where the subpopulation $y$ disappears and we observe the coexistence of subpopulations $x$ and $z$.

\begin{figure}[t]
\begin{center}
\vspace{3mm}
\includegraphics[width=0.7\textwidth,angle=0]{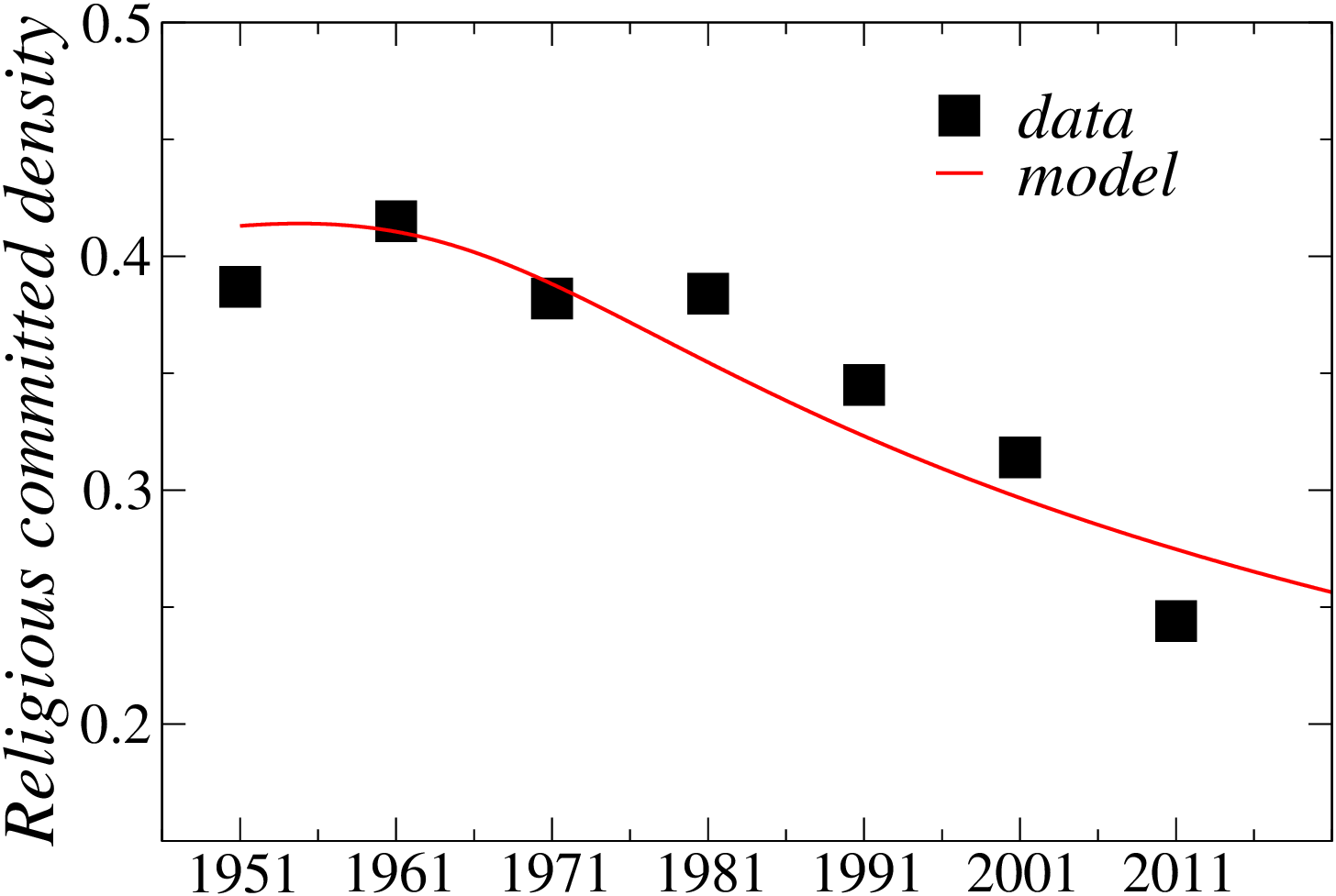}
\end{center}
\caption{(Color online) Comparison between data for the fraction of Northern Irish population who are deemed to be religiously committed (group X, black squares) over time and the model (full red line).  The line is the time evolution of $x(t)$, that was obtained from the numerical integration of Eqs. \eqref{eq1} - \eqref{eq3}. Parameters are $w=0.047$, $q=0.078$, $u=0.32$ and $p=0.043$. Data was obtained from reference \cite{glass}.}
\label{fig4}
\end{figure}

Finally, in order to verify if this simple three-compartment model is able to qualitatively reproduce real data, we compare the model's results with data for religious affiliation in Northern Ireland, available in ref. \cite{glass}. In such reference \cite{glass} we found data of the total number of communicant members of the Presbyterian Church as recorded in church reports, which results in a density of religious committed individuals. Data were collected from 1951 to 2011 in time steps of 10 years, thus in Fig. \ref{fig4} the initial time $t=0$ represents the fraction of religious committed individuals for 1951, $t=1$ represents the fraction of religious committed individuals for 1961, and so on. Since the data is only for the fraction of religious committed individuals, we plot the density $x(t)$ obtained from the model, together with the data. The curve for $x(t)$ was obtained from the numerical integration of Eqs. \eqref{eq1} - \eqref{eq3}. For the numerical results, we considered the parameters are $w=0.047$, $q=0.078$, $u=0.32$ and $p=0.043$, obtained from a least-square fit. One can see in Fig. \ref{fig4} that the numerical results qualitatively correspond to the data. It is important to notice that, for such parameters, the critical points are $p_{c}^{(1)}=0.047$ and $p_{c}^{(2)}\approx 0.0099$. Thus, the fitted parameters represents the population in the region where the three subpopulations $X, Y$ and $Z$ coexist ($p_{c}^{(1)} < p < p_{c}^{(2)}$), which is a realistic region of the model (region II of fig. \ref{fig3}).


%


\subsection{Two-dimensional square lattice}

In this subsection we present some numerical results of Monte Carlo simulations of the model on the top of 2d square lattices. Thus, we built an agent-based formulation of the compartmental model proposed in the last subsection. For this purpose, the algorithm to simulate the model is defined as follows:

\begin{figure}[t]
\begin{center}
\vspace{6mm}
\includegraphics[width=0.45\textwidth,angle=0]{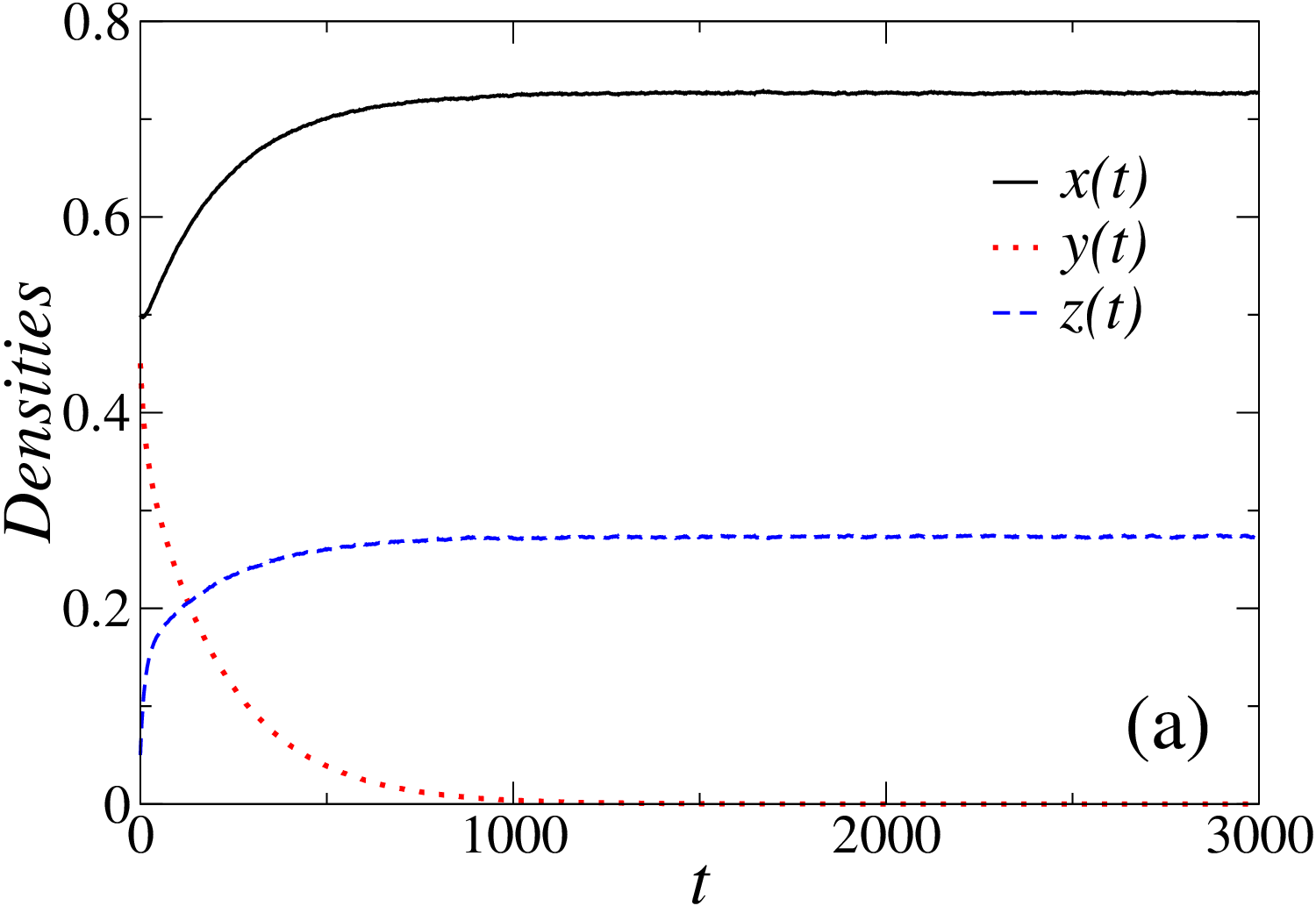}
\hspace{0.3cm}
\includegraphics[width=0.45\textwidth,angle=0]{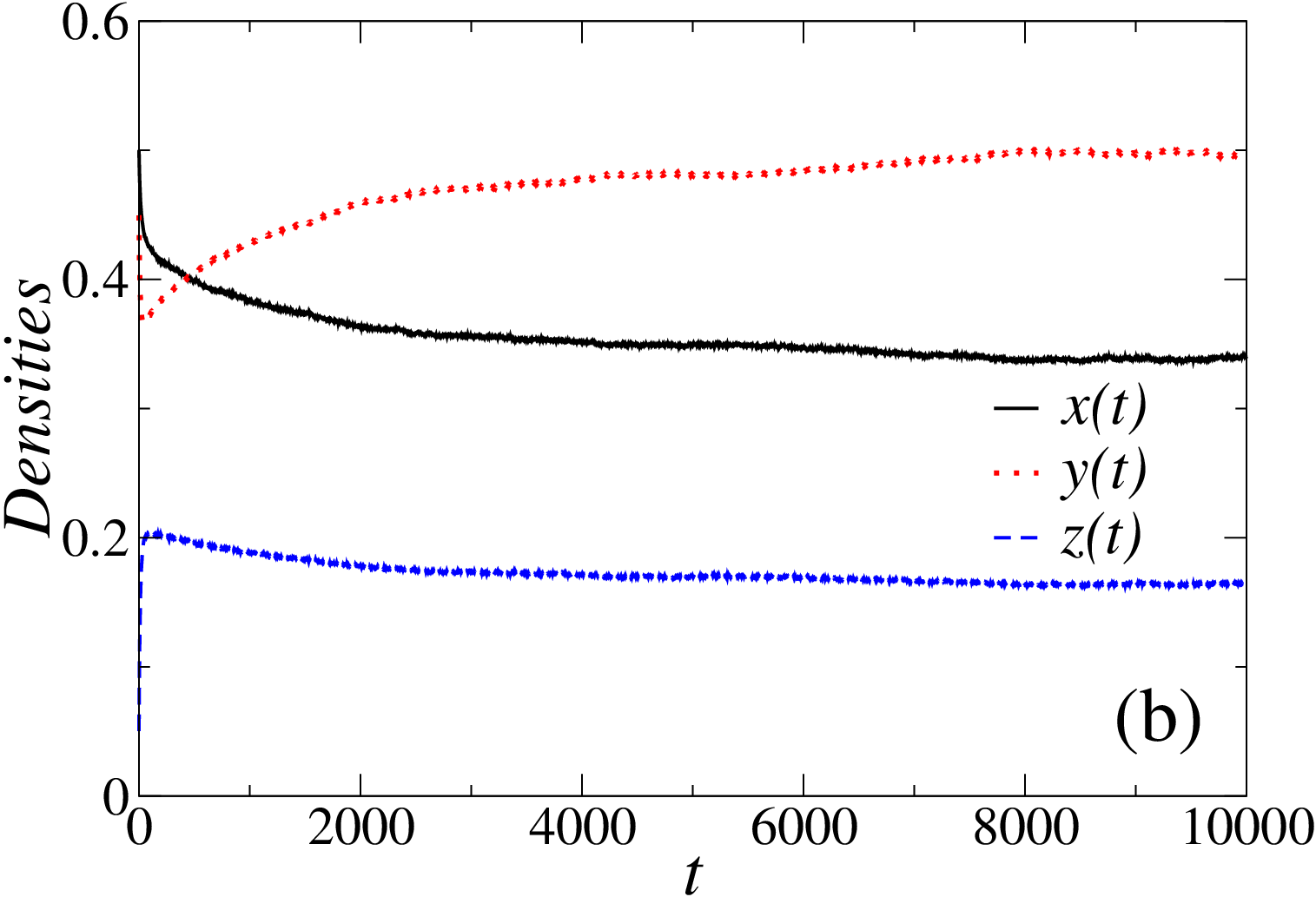}
\\
\vspace{1.0cm}
\includegraphics[width=0.45\textwidth,angle=0]{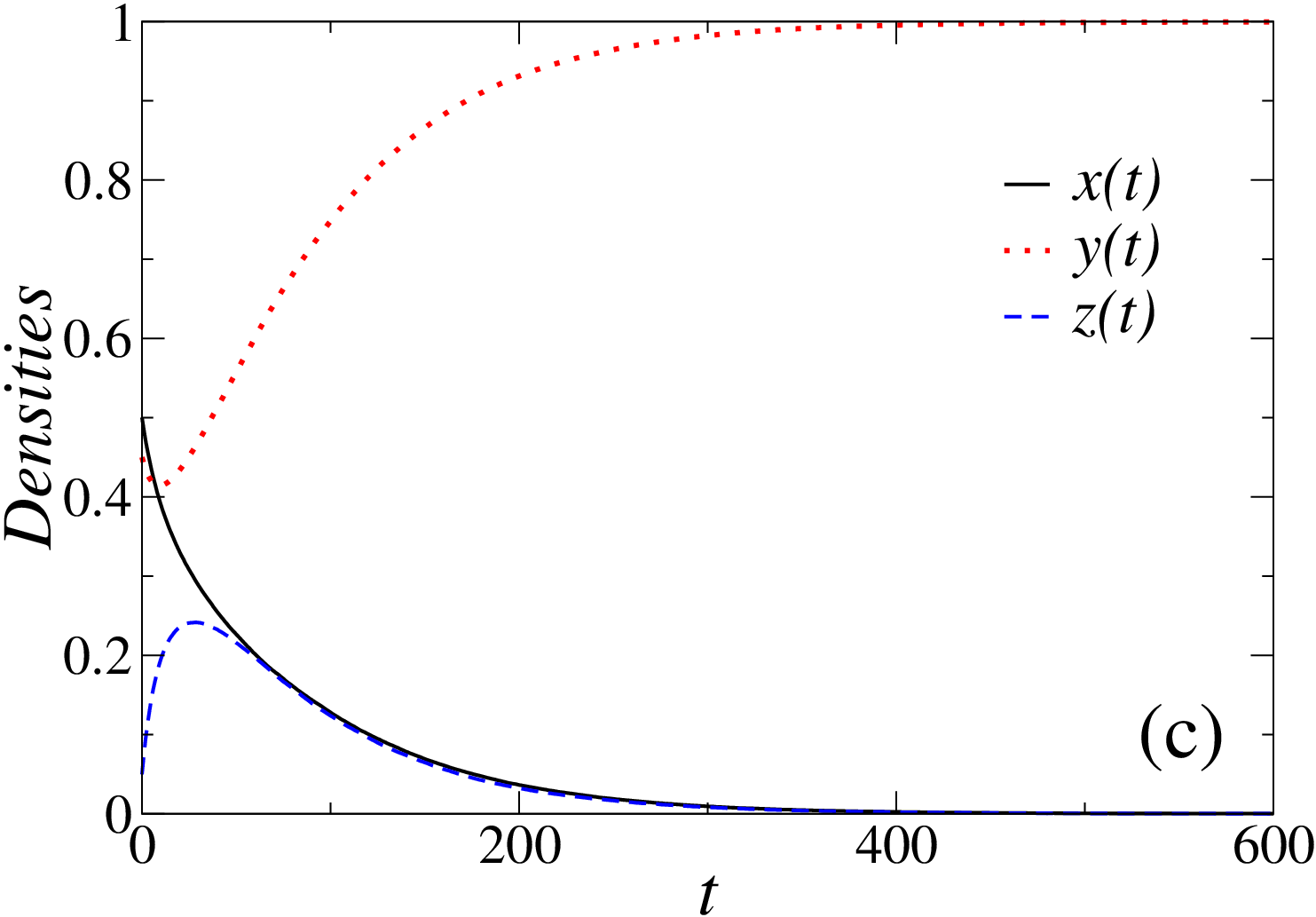}
\hspace{0.3cm}
\includegraphics[width=0.45\textwidth,angle=0]{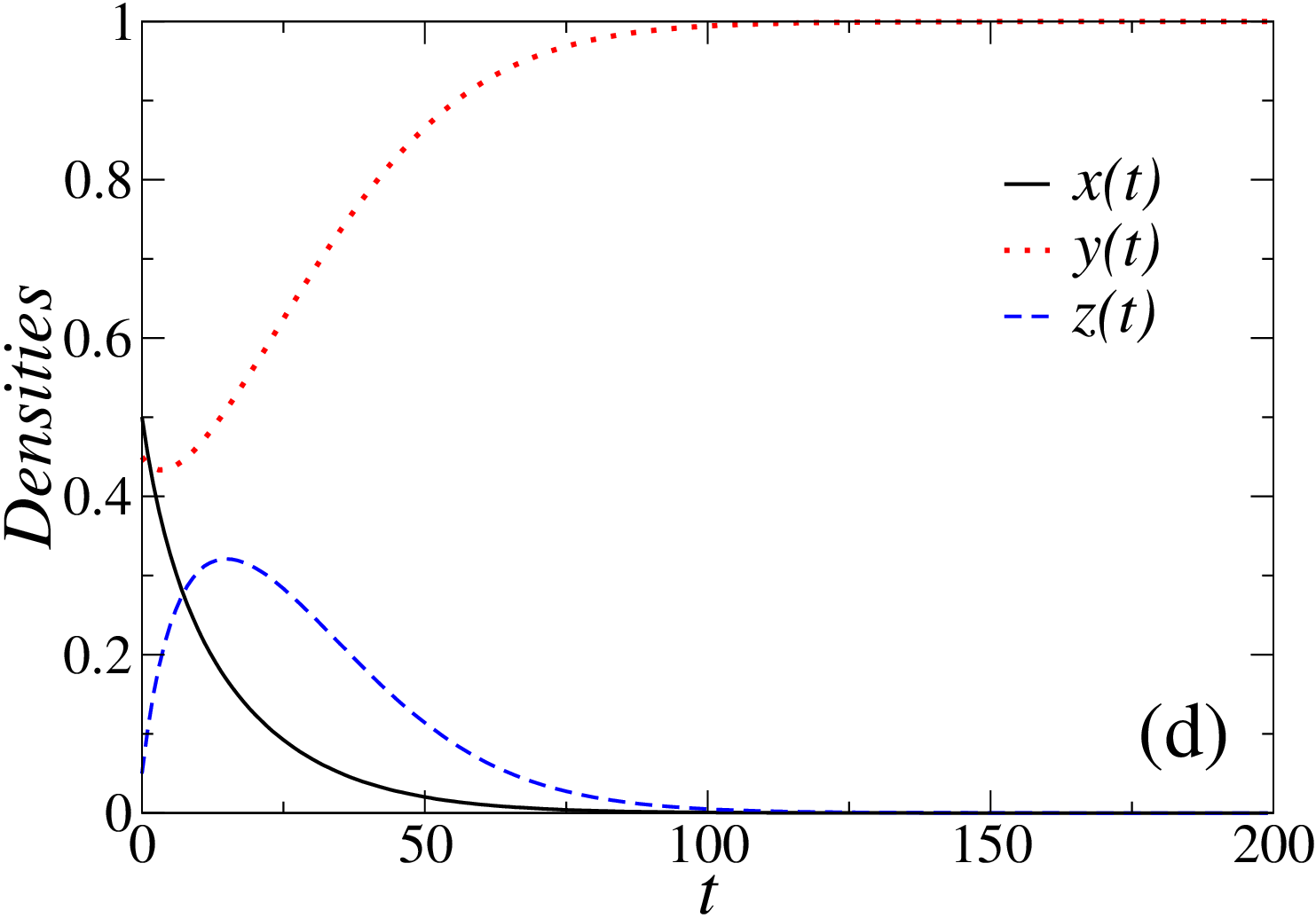}
\end{center}
\caption{(Color online) Time evolution of the three densities of agents $x(t)$, $y(t)$ and $z(t)$ for the model defined on a 2d square lattice with linear size $L=100$. The fixed parameters are $w=0.05$, $q=0.12$ and $u=0.10$, and we varied the parameter $p$: (a) $p=0.030$, (b) $p=0.038$, (c) $p=0.060$, (d) $p=0.120$. Results are averaged over $100$ independent simulations.}
\label{fig5}
\end{figure}

\begin{itemize}

\item we generate a $L$ x $L$ square lattice with a population size $N=L^{2}$ and periodic boundary conditions;

\item given an initial condition $X(0), Y(0)$ and $Z(0)$, we randomly distribute these agents in the lattice sites;

\item at each time step, every lattice site is visited in a sequential order;

\item if a given agent $i$ is in $X$ state, we generate a random number $r$ in the range $[0,1]$. If $r<p$, the agent $i$ changes to state $Z$;

\item if a given agent $i$ is in $Z$ state, we choose at random one of his/her nearest neighbors, say $j$. If such neighbor $j$ is in $X$ state, we generate a random number $r$ in the range $[0,1]$. If $r<q$, the agent $i$ changes to state $X$. On the other hand, if the neighbor $j$ is in $Y$ state, we generate a random number $r$ in the range $[0,1]$. If $r<u$, the agent $i$ changes to state $Y$;

\item if a given agent $i$ is in $Y$ state, we choose at random one of his/her nearest neighbors, say $k$. If such neighbor $k$ is in $X$ state, we generate a random number $r$ in the range $[0,1]$. If $r<w$, the agent $i$ changes to state $X$.
\end{itemize}

One time step is defined by the visit of all lattice sites. The agents' states were updated synchronously, i.e., we considered parallel updating, as it is standard in probabilistic cellular automata in order to avoid correlations between consecutive steps \cite{mf_keom,roos}. In addition, all results are averaged over $100$ independent simulations.

Considering the same parameters of the previous subsection, namely $w=0.05$, $q=0.12$ and $u=0.10$, and varying the parameter $p$, we exhibit in Fig. \ref{fig5} the time evolution of the system on a square lattice of linear size $L=100$. We can see a similar behavior observed in the fully-connected case, i.e., we can have the coexistence of $x$ and $z$ populations in the steady states (panel (a)), the coexistence of the three subpopulations (panel (b)) and absorbing states with $y=1$, $x=z=0$ (panels (c) and (d)). However, the time to reach the steady states is longer than in the mean-field case, as it is usual in agent-based models defined on lattices \cite{pmco}.

\begin{figure}[t]
\begin{center}
\vspace{3mm}
\includegraphics[width=0.7\textwidth,angle=0]{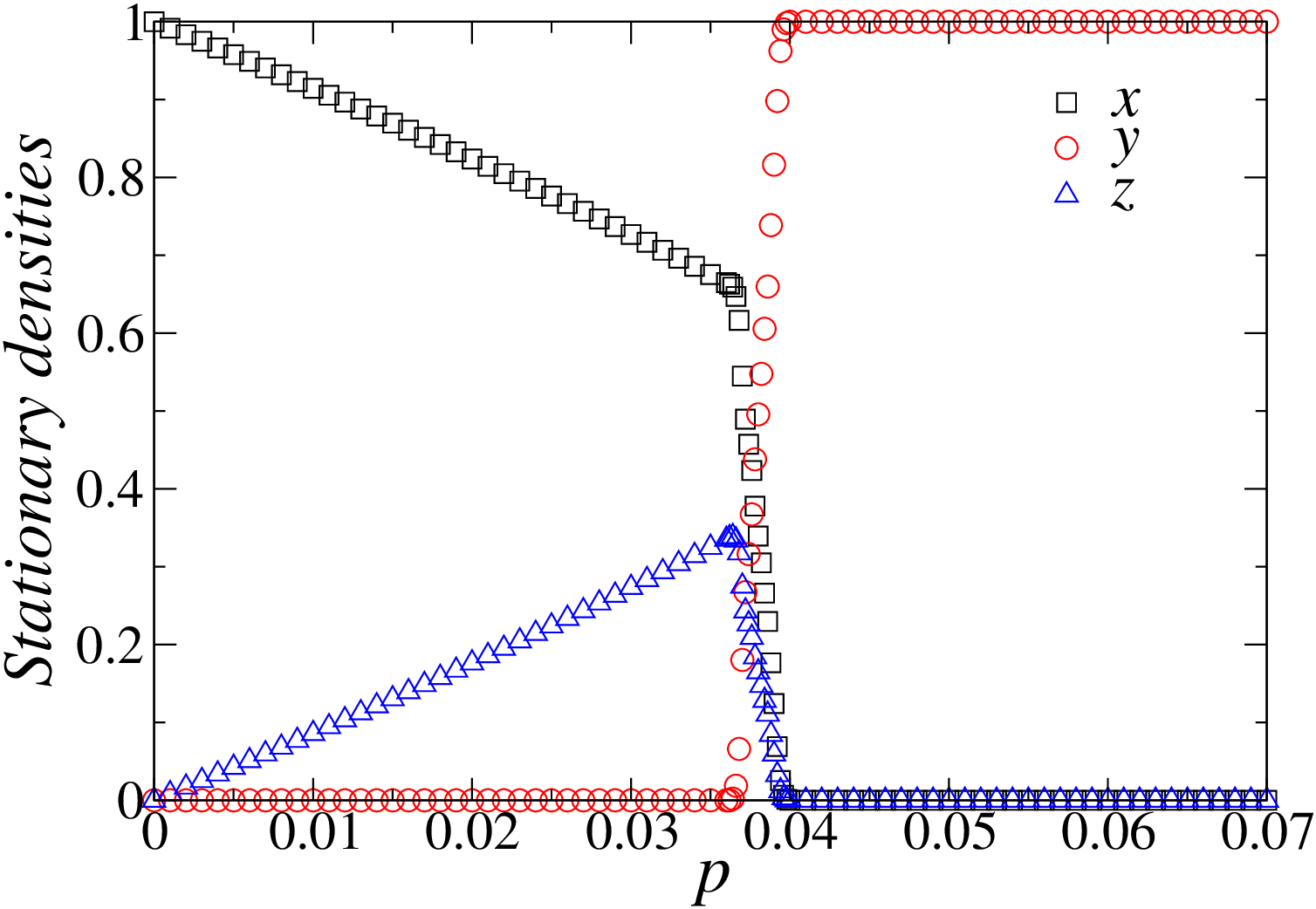}
\end{center}
\caption{(Color online) Stationary densities $x$, $y$ and $z$ as functions of the probability $p$ for $w=0.05$, $q=0.12$ and $u=0.10$ for the model defined on a 2d square lattice with linear size $L=100$. The absorbing state where $x=z=0$ and $y=1$ is observed for $p>\approx 0.04$. Results are averaged over $100$ independent simulations.} 
\label{fig6}
\end{figure}

We can also compute the stationary densities $x, y$ and $z$ to compare with the mean-field case. In Fig. \ref{fig6} we exhibit the stationary values of the three subpopulations as function of the probability $p$ for $w=0.05$, $q=0.12$ and $u=0.10$, considering a square lattice with linear size $L=100$. We observe similar behaviors for such quantities in comparison with the mean-field calculations presented in section 3.1. However, as it is usual in lattice models, we observe a distinct location of the critical points $p_{c}^{(1)}$ and $p_{c}^{(2)}$ in comparison with the mean-field predictions \cite{dickman}. For the same set of parameters, looking for the data for the 2d case we have $p_{c}^{(1)}\approx 0.0364$ and $p_{c}^{(2)}\approx 0.0402$, whereas in the previous subsection we observed $p_{c}^{(1)}=0.04$ and $p_{c}^{(2)}=0.05$. The differences can be pointed to the correlations generated by the lattice structure, that are absent in the fully-connected case.

\begin{figure}[t]
\begin{center}
\vspace{6mm}
\includegraphics[width=0.45\textwidth,angle=0]{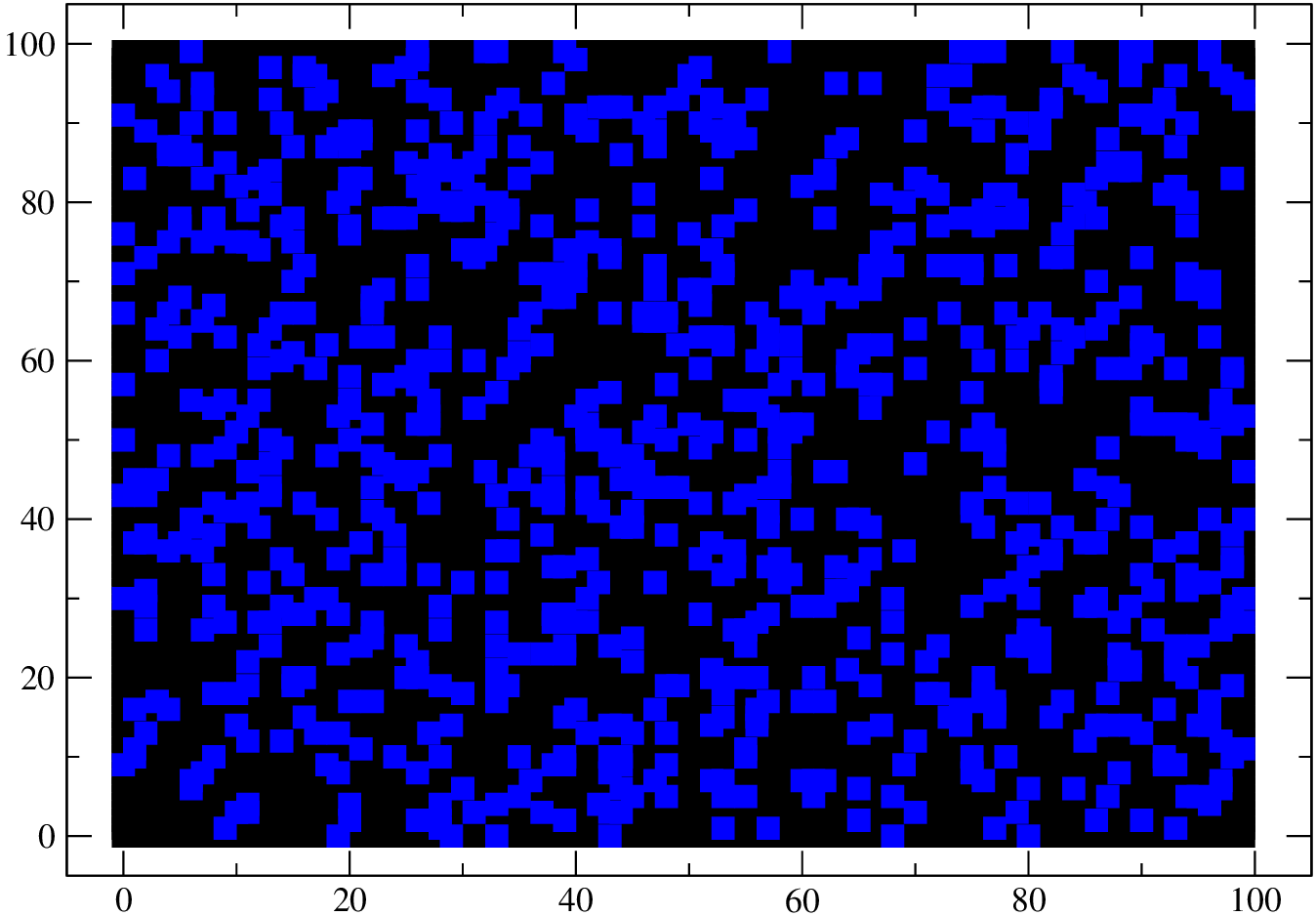}
\hspace{0.3cm}
\includegraphics[width=0.45\textwidth,angle=0]{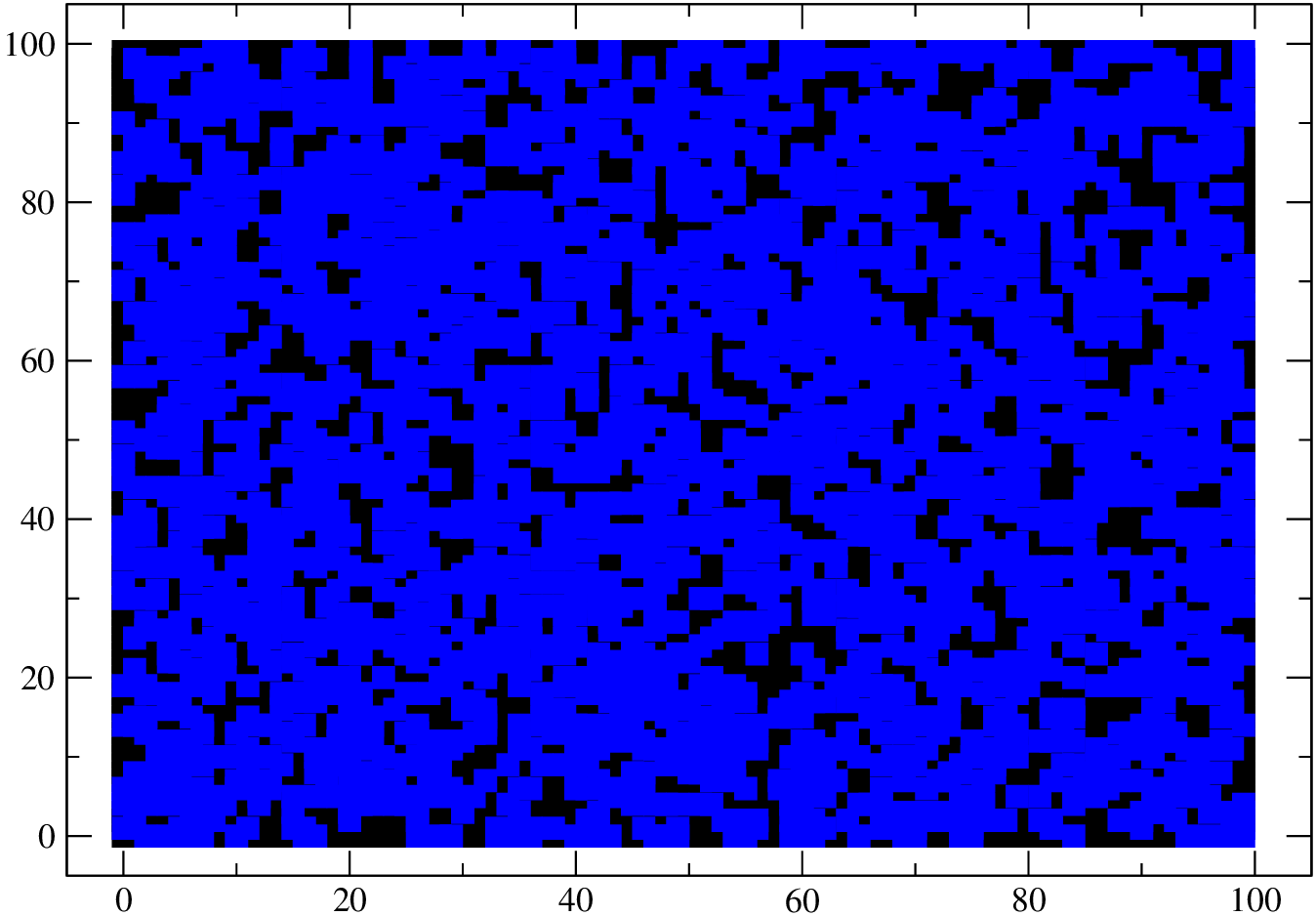}
\\
\vspace{1.0cm}
\includegraphics[width=0.45\textwidth,angle=0]{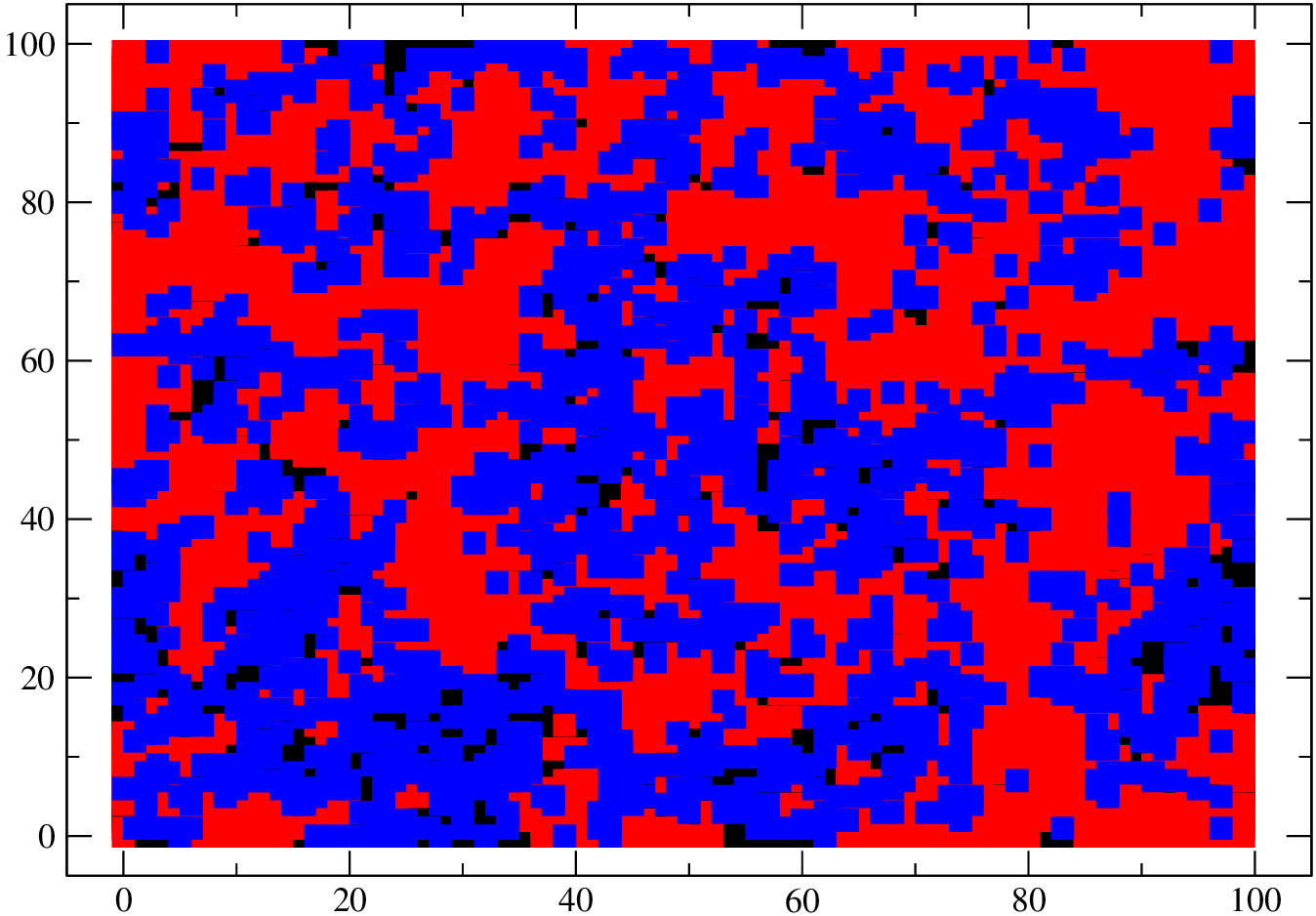}
\hspace{0.3cm}
\includegraphics[width=0.45\textwidth,angle=0]{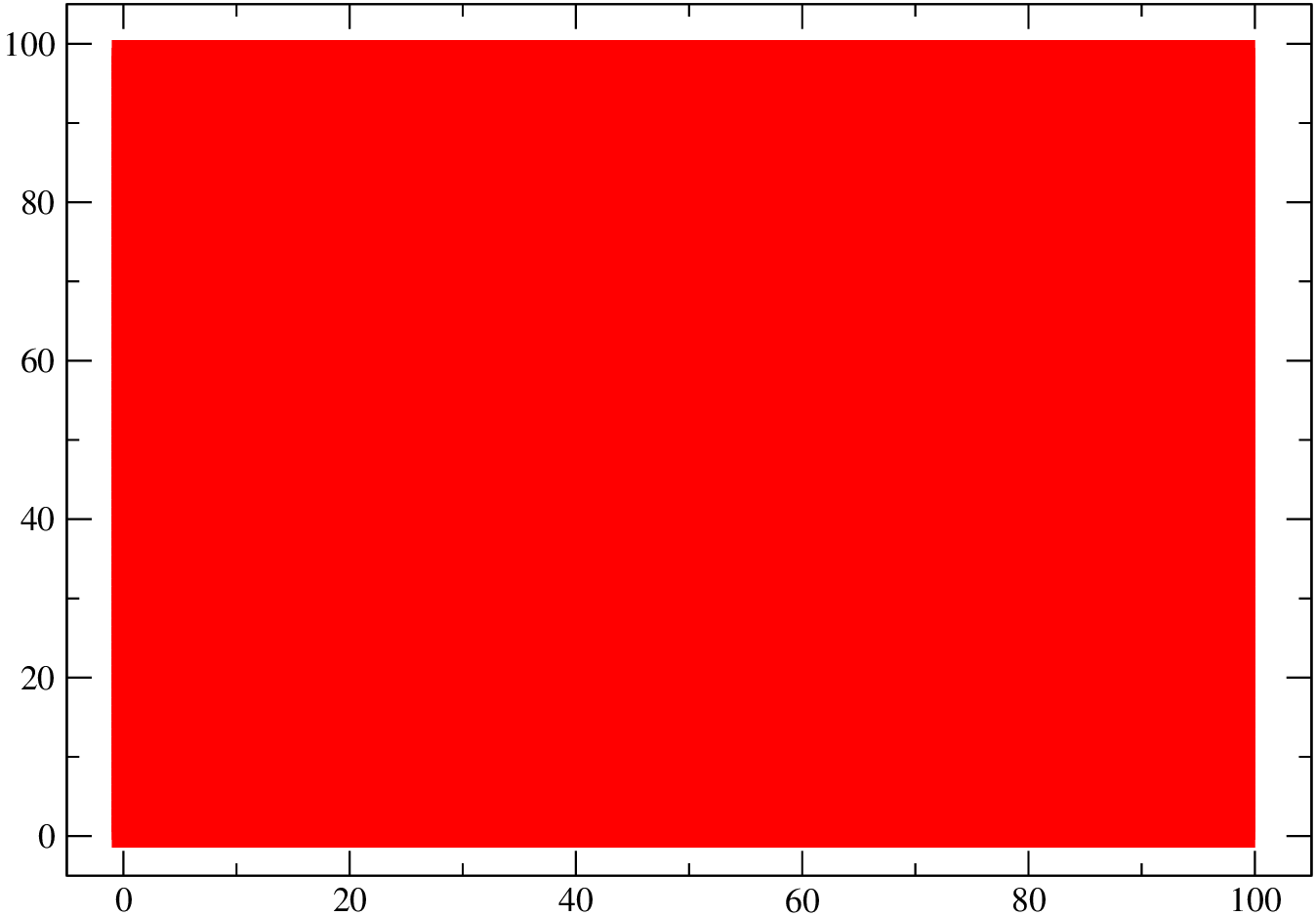}
\end{center}
\caption{(Color online) Snapshots of the stationary states of the model on a square lattice of linear size $L=100$. The distinct colors represent the subpopulations $x$ (black), $y$ (red) and $z$ (blue). The fixed parameters are $w=0.05$, $q=0.12$ and $u=0.10$, and we varied the parameter $p$: (a) $p=0.010$, (b) $p=0.030$, (c) $p=0.038$, (d) $p=0.050$.}
\label{fig7}
\end{figure}

For better visualization of the lattice and the subpopulations, we plot in Fig. \ref{fig7} some snapshots of the stationary states of the model. We considered the same lattice size used before, namely $L=100$, the same fixed parameters $w=0.05$, $q=0.12$ and $u=0.10$, and we plot four distinct values of $p$. The distinct colors represent the subpopulations $x$ (black), $y$ (red) and $z$ (blue). For low values of $p$ like $p=0.010$, we observe the dominance of the $x$ population (upper panel, left side), and no $y$ population. If we increase $p$ for $p=0.030$ we observe an increase of the $z$ population and the decrease of the $x$ population, and the $y$ population remains absent (upper panel, right side). For $p=0.038$ we are in the region where the three subpopulations coexist (lower panel, left side). Finally, for sufficient high values of $p$ like $p=0.050$ we observe that only the $y$ population survives in the long-time limit (lower panel, right side), which represents the previous mentioned absorbing state of the model.

\section{Model II: Presence of inflexible individuals}

In this section we consider the presence of inflexible individuals in the population. Such individuals will be included in the subpopulations of religious committed individuals ($\textbf{X}$) and not religious ones ($\textbf{Y}$). Inflexible, or zealot, individuals are the ones that never change their status - in this case, either $\textbf{X}$ or $\textbf{Y}$ individuals will remain in their states during all the dynamics. This is analogous to introducing quenched disorder in the system, considerably altering its critical behavior.

In this case, as considered recently for the dynamics of alcohol consumption \cite{drink}, we consider that we have a certain number $X_{I}$ of inflexible agents related to religious commited individuals, i.e., individuals that are committed with a given religion and will always devoted to such reiligion, i.e., they will keep this behavior during all their lives. In addition, we also considered a certain number $Y_{I}$ of inflexible agents related to not religious individuals, i.e., individuals that do not follow any religion and do not intend to do it. Thus, the total number of religious committed individuals are given by $X_{total}=X + X_{I}$ and the total number of not religious individuals are given by $Y_{total}=Y+Y_I$, where $X$ and $Y$ denote the noninflexible individuals of each class. Since we are not considering inflexible individuals for the religious noncommitted individuals $Z$, we have simply $Z_{total}=Z$.

For this model, the transitions among compartments are the following:

\begin{itemize}
\item $X \stackrel{p}{\rightarrow} Z$: a noninflexible religious committed agent (X) becomes a religious noncommitted individual (Z) with probability $p$;

\item $Z \stackrel{q}{\rightarrow} X$: a religious noncommitted agent (Z) becomes a religious committed individual (X) with probability $q$ if he/she is in contact with religious committed individuals (X and $X_{I}$);

\item $Z \stackrel{u}{\rightarrow} Y$: a religious noncommitted agent (Z) becomes a not religious affiliated individual (Y) with probability $u$ if he/she is in contact with not religious affiliated individuals (Y and $Y_{I}$);

\item $Y \stackrel{w}{\rightarrow} X$: a noninflexible not religious affiliated agent (Y) becomes a religious committed individual (X) with probability $w$ if he/she is in contact with religious committed individuals (X and $X_{I}$);

\end{itemize}

For such a case, the normalization condition becomes
\begin{equation} \label{eq9}
x(t) + y(t) + z(t) + x_I + y_I = 1 ~,
\end{equation}
\noindent
where $x_I = X_I/N$ and $y_I=Y_I/N$ are the inflexible/zealot subpopulations in the compartments $X$ and $Y$, respectively, that are constant in time.

In the next section we discuss the impact of such zealots in the critical behavior of the model.


\section{Results: Model II}

\subsection{Fully-connected population}

Considering the above-mentioned rules for the model in the presence of inflexible individuals and a population defined on a fully-connected network, we can write the following mean-field equations for the compartments' evolution:
\begin{eqnarray} \label{eq10}
\frac{dx(t)}{dt} & = & -p\,x(t) + q\,[x(t)+x_I]\,z(t) + w\,[x(t)+x_I]\,y(t) \\  \label{eq11}
\frac{dy(t)}{dt} & = & -w\,[x(t)+x_I]\,y(t) + u\,[y(t)+y_I]\,z(t) \\ \label{eq12}
\frac{dz(t)}{dt} & = & -u\,[y(t)+y_I]\,z(t) - q\,[x(t)+x_I]\,z(t) + p\,x(t) 
\end{eqnarray}

\begin{figure}[t]
\begin{center}
\vspace{6mm}
\includegraphics[width=0.45\textwidth,angle=0]{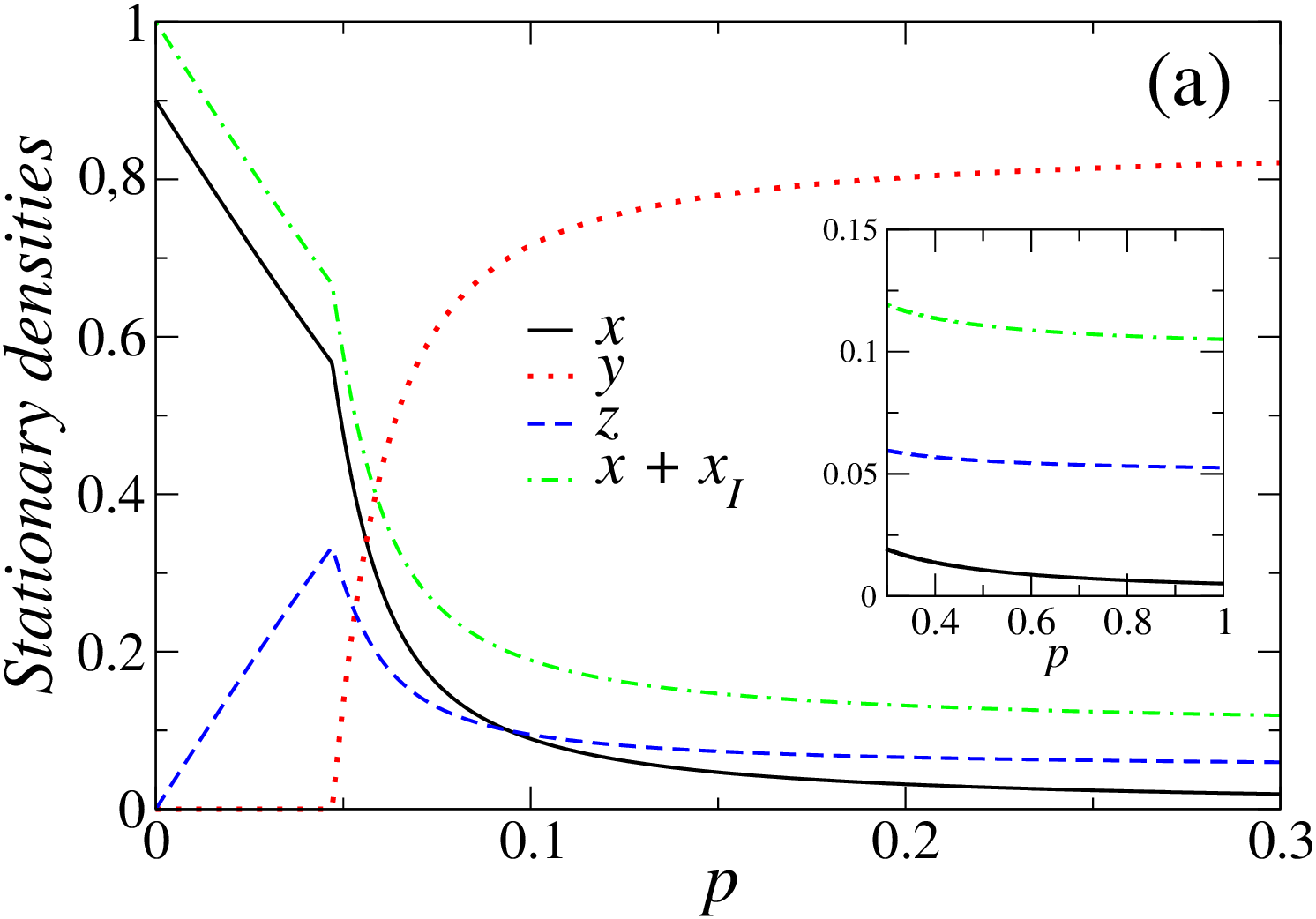}
\hspace{0.3cm}
\includegraphics[width=0.45\textwidth,angle=0]{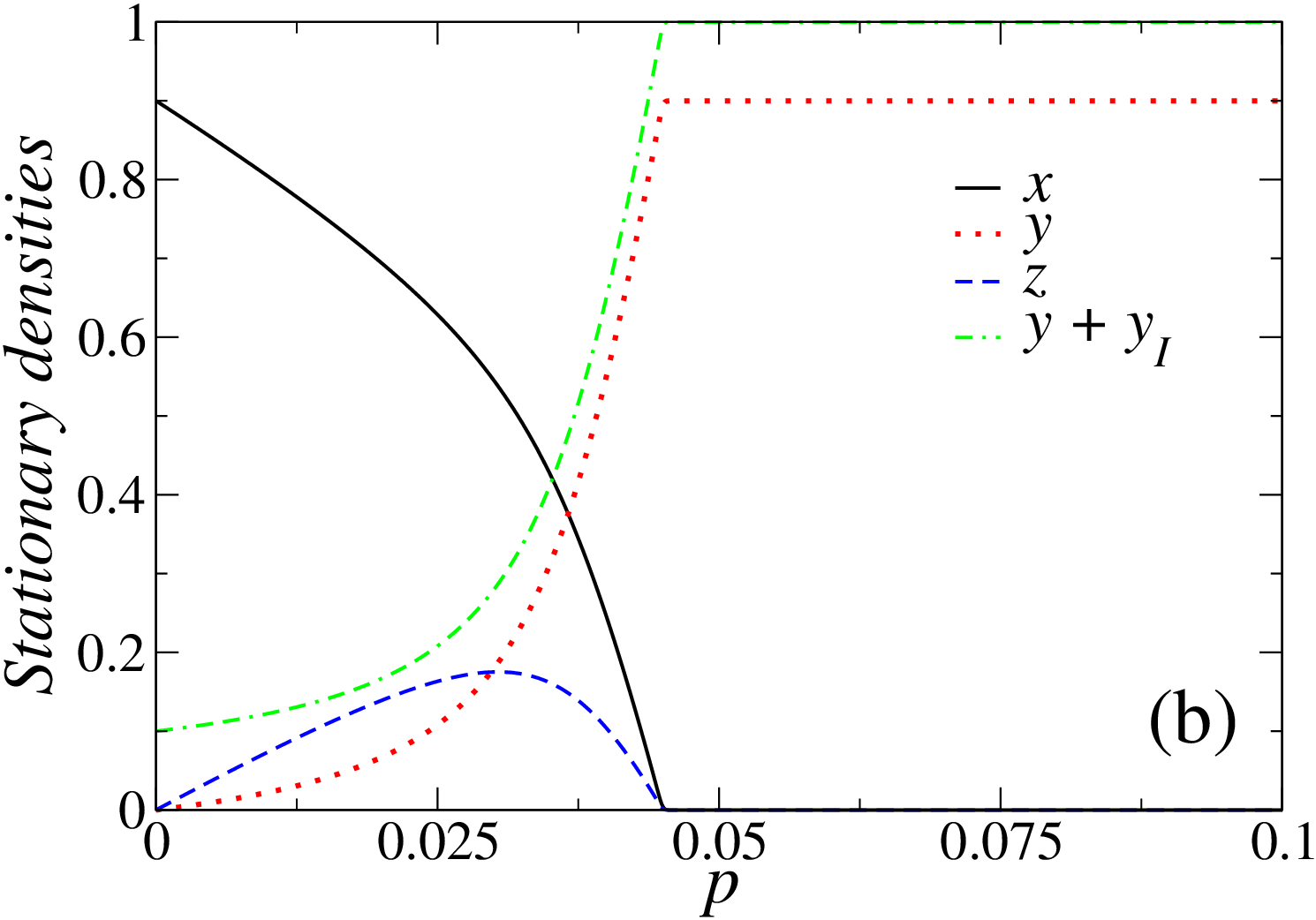}
\\
\vspace{1.0cm}
\includegraphics[width=0.45\textwidth,angle=0]{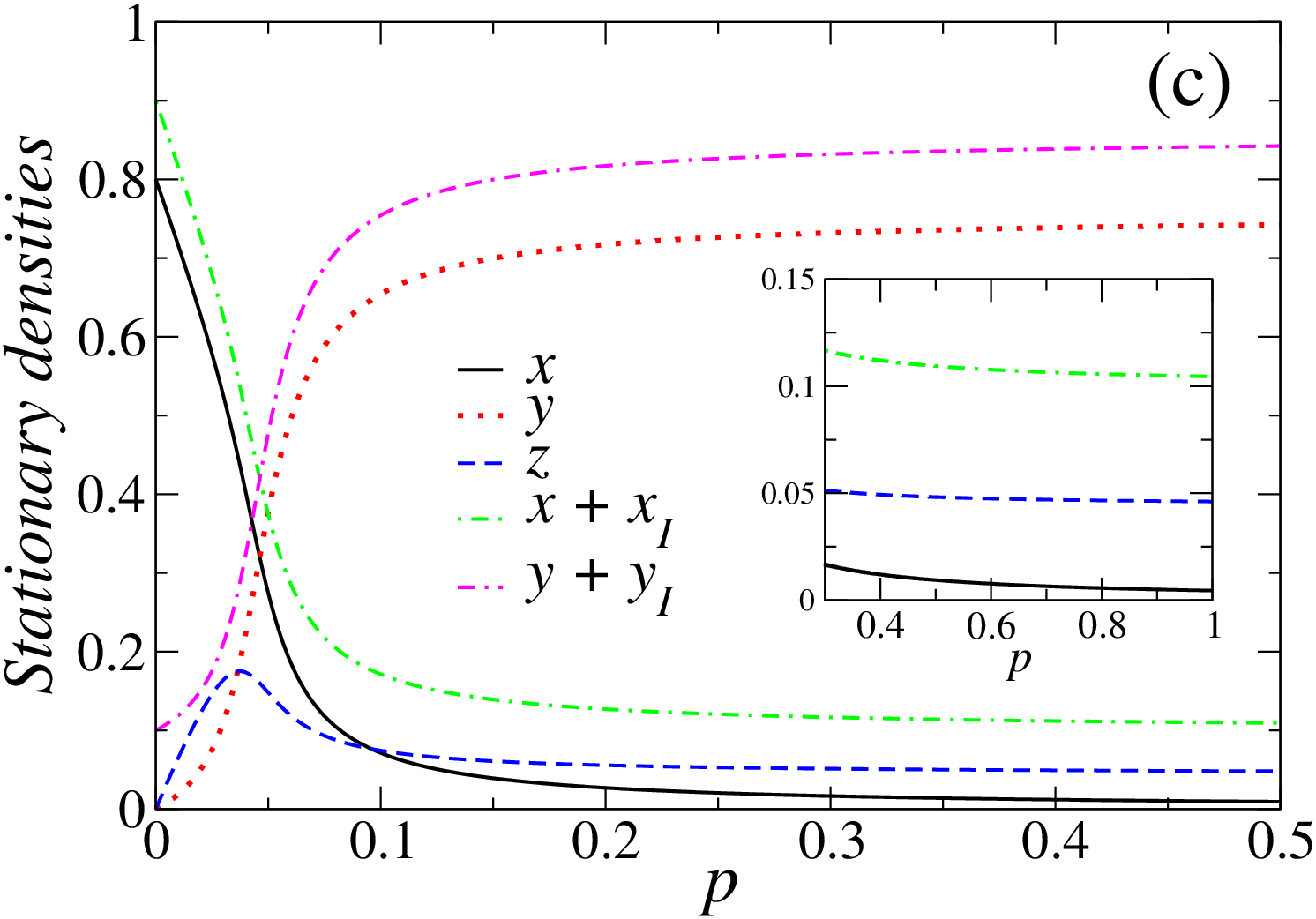}
\end{center}
\caption{(Color online) Stationary densities of the subpopulations as functions of the probability $p$ for $w=0.05$, $q=0.12$ and $u=0.10$, for the mean-field formulation of the model in the presence of inflexible individuals. The lines were obtained from the numerical integration of Eqs. \eqref{eq10} - \eqref{eq12}. The densities of inflexible/zealots are: (a) $x_{I}=0.1, y_{I}=0.0$, (b) $x_{I}=0.0, y_{I}=0.1$ and (c) $x_{I}=0.1, y_{I}=0.1$.}
\label{fig8}
\end{figure}

Since we are interested in the critical behavior of the model, we present in Fig. \ref{fig8} only the stationary densities of the subpopulations as functions of $p$. The fixed parameters are the same used in the remaining of the work, i.e., $w=0.05$, $q=0.12$ and $u=0.10$. To analyze the impact of the distinct classes of inflexible individuals $x_{I}$ and $y_{I}$, we plot in Fig. \ref{fig8} (a) the stationary values only in the presence of inflexible agents in the class of religious committed individuals. In this case, we exhibit results for $x_{I}=0.1$ and $y_{I}=0.0$. We can see the phase where $y=0$ in the steady states persists for low values of $p$, and we observe there is a critical point associated with the transition from such phase where $y=0$ to $y \neq 0$. The second phase transition (active-absorbing transition) is eliminated in the presence of inflexible agents in the class of religious committed individuals, i.e., we can see that the subpopulations of religious committed individuals and not religious ones will not disappear of the population for long times, for all values of $p$ (see the inset of Fig. \ref{fig8} (a)). In other words, the absorbing state is suppressed in the presence of $X_{I}$ inflexible individuals.

In Fig. \ref{fig8} (b) we exhibit the stationary densities only in the presence of inflexible agents in the class of not religious affiliated individuals. In this case, we exhibit results for $x_{I}=0.0$ and $y_{I}=0.1$. We can see the collective state where the class of not religious affiliated individuals disappear of the population is eliminated. However, we observe the persistence of the active-absorbing phase transition. However, such phase is composed now by $Y$ and $Y_{I}$ individuals, and the remaining subpopulations $X$ and $Z$ disappear.

Finally, the join effect of the inflexible individuals in the population are shown in Fig. \ref{fig8} (c), where we show results for $x_{I}=0.1$ and $y_{I}=0.1$. We see that both phase transitions are now suppressed, and we have the coexistence of all the subpopulations $X, X_{I}, Y, Y_{I}$ and $Z$ for all values of $p>0$. Since the coexistence of the distinct classes of individuals is a realistic scenario, we can argue that the presence of inflexible individuals in both classes (religious committed agents and not religious affiliated ones) makes the model more realistic.


\subsection{Two-dimensional square lattice}

In this subsection we present some numerical results of Monte Carlo simulations of the model with inflexible agents on the top of 2d square lattices. Thus, we built an agent-based formulation of the compartmental model proposed in the last subsection. We will keep the nomenclature of the last subsection, i.e., the noninflexible population in the three classes is given by $X, Y$ and $Z$, and the inflexible population is given by $X_{I}$ and $Y_{I}$. Thus, the algorithm to simulate the model is defined as follows:

\begin{itemize}

\item we generate a $L$ x $L$ square lattice with a population size $N=L^{2}$ and periodic boundary conditions;

\item we choose at random $X_{I}$ ($Y_{I}$) lattice sites and freeze their opinions in the religious committed individuals (not religious affiliated individuals), in such a way that such individuals will never change opinion during all the dynamics;

\item the remaining of individuals in the population are distributed in the noninflexbile states $X, Y$ and $Z$ with a given initial condition $X(0), Y(0)$ and $Z(0)$, and the are randomly distributed over the lattice;

\item at each time step, every lattice site is visited in a sequential order;

\item if a given agent $i$ is in $X$ state, we generate a random number $r$ in the range $[0,1]$. If $r<p$, the agent $i$ changes to state $Z$;

\item if a given agent $i$ is in $Z$ state, we choose at random one of his/her nearest neighbors, say $j$. If such neighbor $j$ is in $X$ or $X_{I}$ states, we generate a random number $r$ in the range $[0,1]$. If $r<q$, the agent $i$ changes to state $X$. On the other hand, if the neighbor $j$ is in $Y$ or $Y_{I}$ states, we generate a random number $r$ in the range $[0,1]$. If $r<u$, the agent $i$ changes to state $Y$;

\item if a given agent $i$ is in $Y$ state, we choose at random one of his/her nearest neighbors, say $k$. If such neighbor $k$ is in $X$ or $X_{I}$ states, we generate a random number $r$ in the range $[0,1]$. If $r<w$, the agent $i$ changes to state $X$.
\end{itemize}

As in section 3.2, one time step is defined by the visit of all lattice sites, and the agents' states were updated syncronously. In addition, all results are averaged over $100$ independent simulations.

\begin{figure}[t]
\begin{center}
\vspace{3mm}
\includegraphics[width=0.45\textwidth,angle=0]{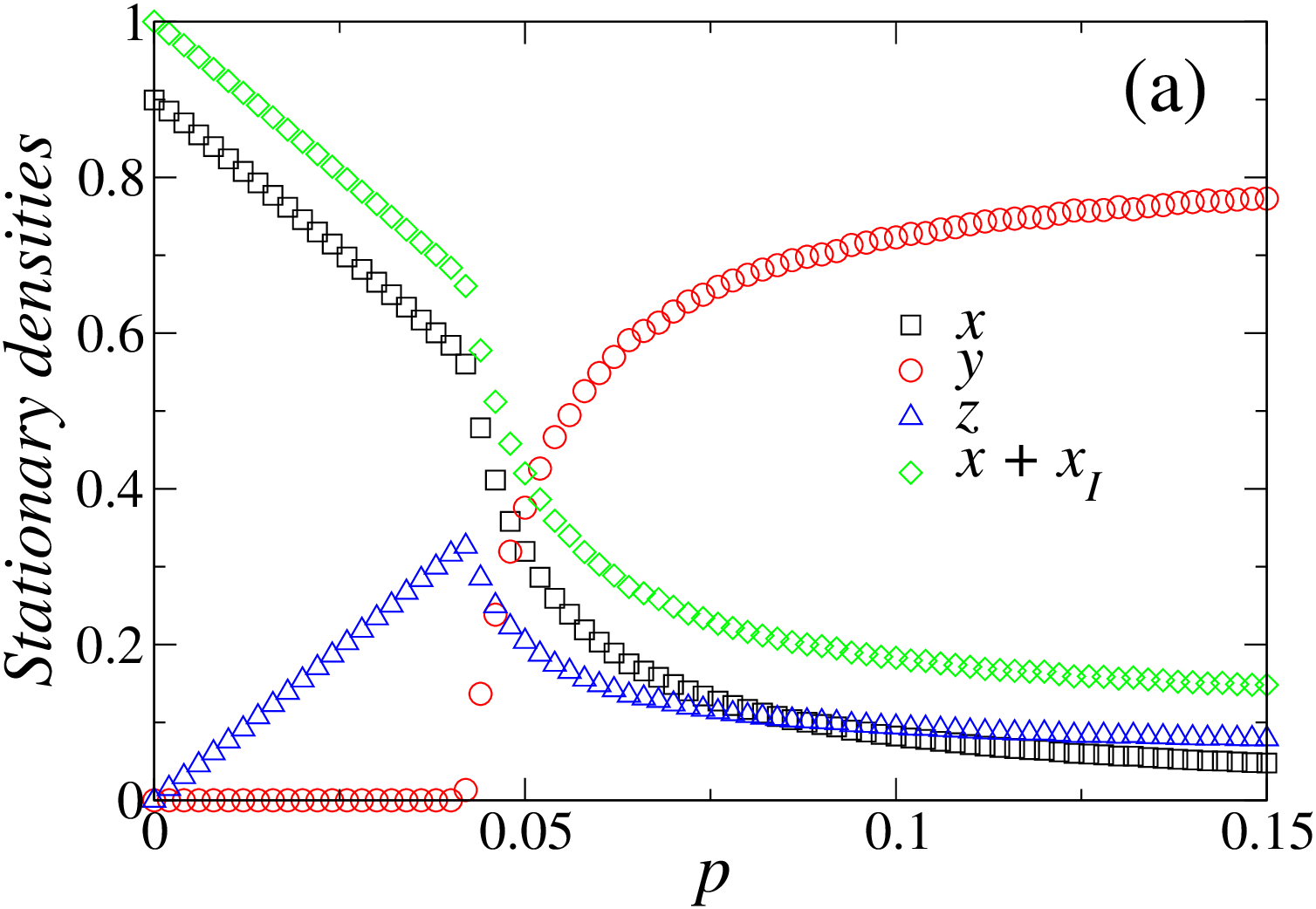}
\hspace{0.3cm}
\includegraphics[width=0.45\textwidth,angle=0]{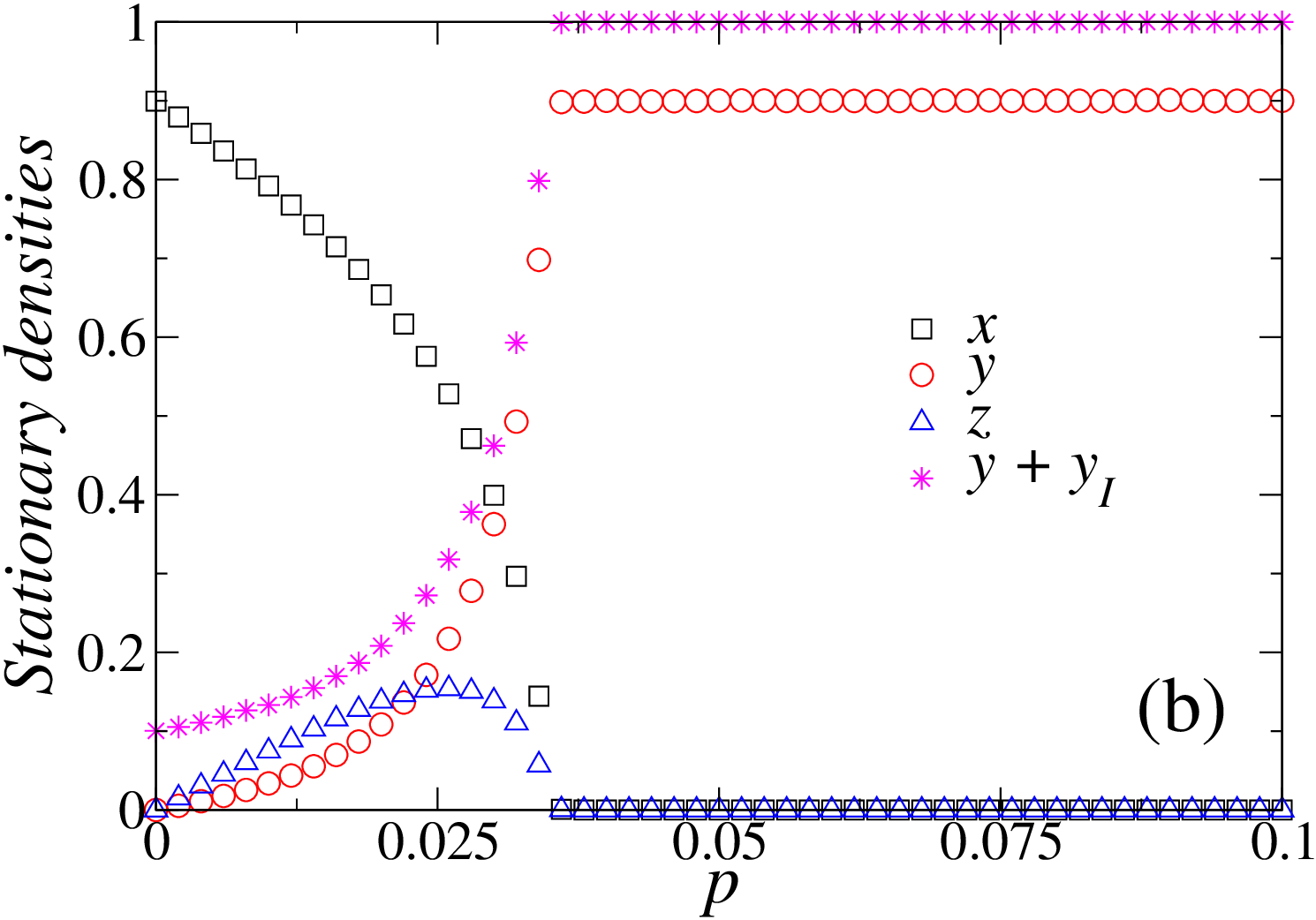}
\\
\vspace{1.0cm}
\includegraphics[width=0.45\textwidth,angle=0]{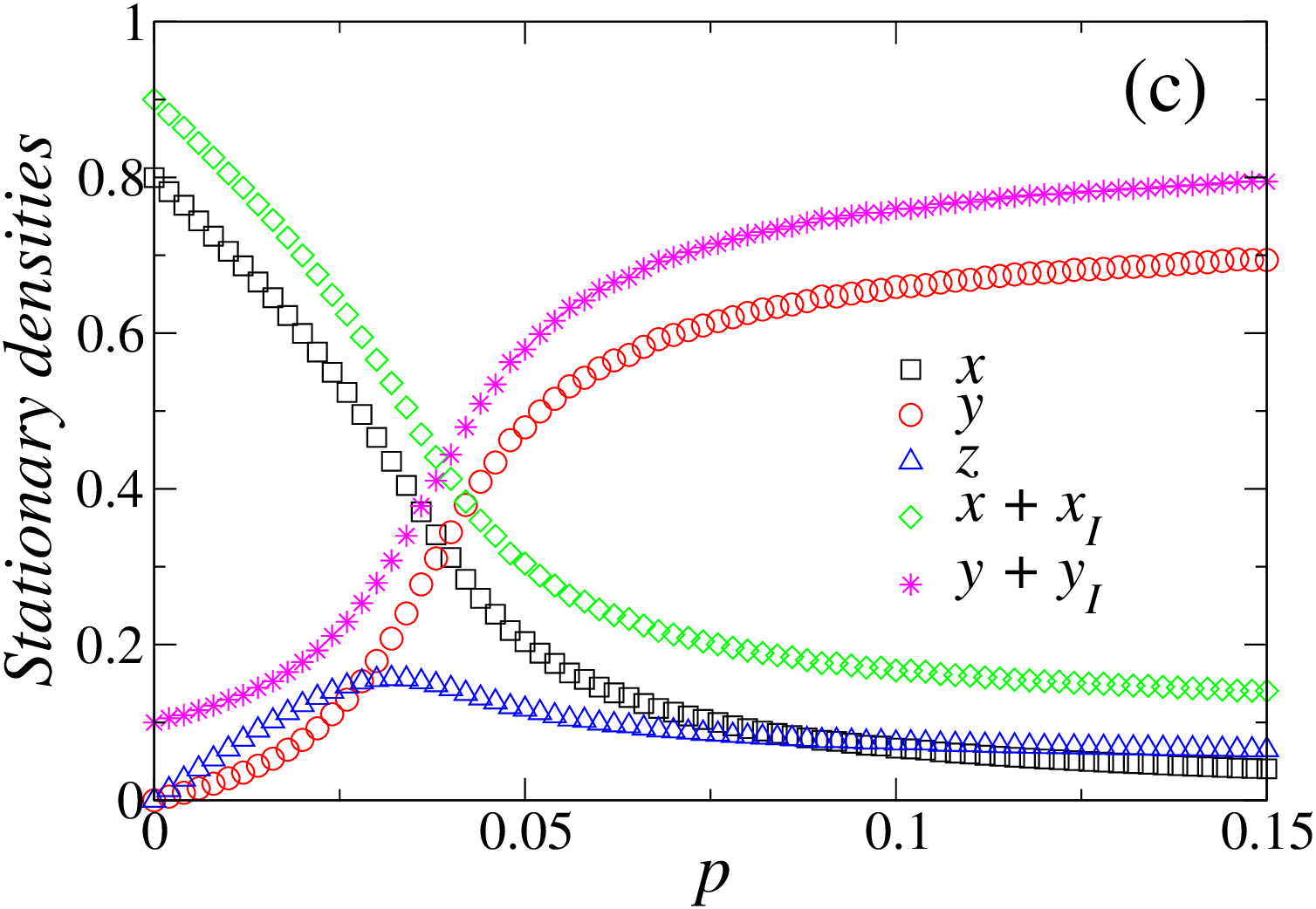}
\end{center}
\caption{(Color online) Stationary densities of the subpopulations as functions of the probability $p$ for $w=0.05$, $q=0.12$ and $u=0.10$ for the model in the presence of inflexible individuals in a square lattice of linear size $L=100$. The densities of inflexible/zealots are: (a) $x_{I}=0.1, y_{I}=0.0$, (b) $x_{I}=0.0, y_{I}=0.1$ and (c) $x_{I}=0.1, y_{I}=0.1$. Results are averaged over $100$ independent simulations.}
\label{fig9}
\end{figure}

Considering the same parameters of the previous subsection, namely $w=0.05$, $q=0.12$ and $u=0.10$, and varying the parameter $p$, we exhibit in Fig. \ref{fig9} the stationary densities of the subpopulations for: $x_{I}=0.1$ and $y_I=0.0$ (panel (a)), $x_{I}=0.0$ and $y_I=0.1$ (panel (b)) and $x_{I}=0.1$ and $y_I=0.1$ (panel (c)). The behavior is similar to the observed in the mean-field case, but again the critical points, when they exist, are located in smaller values of $p$ in comparison with the mean-field case. The suppression of the absorbing phase is also observed for the model in the presence of only $X_{I}$ inflexible agents. In addition, the coexistence of two subpopulations does not occur anymore, as in the model defined in a fully-connected population. Finally, the model in the presence of both $X_{I}$ and $Y_{I}$ inflexible individuals shows the coexistence of all subpopulations for all values of $p$, denoting a more realistic situation.

\section{Final Remarks}   

In this work we study a simple model for the dynamics of religious affiliation. Fot this purpose, we built a three-state compartmental model, and the transitions among such compartments are ruled by probabilities. The three states are: religious committed individuals ($\textbf{X}$), religious noncommitted individuals ($\textbf{Z}$) and not religious affiliated individuals ($\textbf{Y}$). The transitions among the states are ruled by probabilities.

First of all, we study the model on a fully-connected population. Thus, we write the mean-field equations for the evolutions of the three subpopulations. Some analytical results were obtained for the stationary states of the model. Such solutions, together with the numerical integration of the model's equations, reveal that there are three distinct collective states or phases in the stationary states of the model, depending on the range of parameters: (I) a phase where only the $\textbf{Y}$ subpopulation survives; (II) a region where the subpopulations $\textbf{X}$ and $\textbf{Z}$ survive and the $\textbf{Y}$ population disappers; (III) a region where the three subpopulations coexist. Region (I) represents a region of theoretical interest for nonequilibrium Statistical Physics, since it represents an absorbing state, where the system is frozen and no transitions among compartments occur anymore. Region (II) represents a realistic representation of a society, where there are religious individuals (some of them committed with the religion, and some others no committed) and not religious ones. The transition from region II to region I represents an active-absorbing nonequilibrium phase transition that appears to be in the directed percolation universality class. The model is also capable to qualitatively reproduce data for religious affiliation in Northern Ireland. In addition, we also presented in this work results of agent-based Monte Carlo simulations of the model on two-dimensional square lattices. The results show a similar behavior in comparison with the mean-field case, but the critical points are localized for lower values of the control parameter of the phase transition. In addition, the times to the system achieves stationary states are higher in comparison with the mean-field case.

After such initial analysis, we considered the presence of inflexible individuals in the population. For this purpose, we included inflexible/zealot individuals in the subpopulations of religious committed individuals (fixed density $x_{I}$) and not religious ones (fixed density $y_{I}$). Inflexible, or zealot, individuals are the ones that never change their status - in this case, either $\textbf{X}$ or $\textbf{Y}$ individuals will remain in their states during all the dynamics. The densities of such inflexible agents are fixed in time, representing quenched disorder in the language of magnetic systems \cite{book}. For the practical point of view it is a realistic situation that can model the presence of radical individuals concerning religions, and the presence of inflexible individuals can dramastically change the dynamics and the critical behavior of social models \cite{rmp}. We verified that the presence of inflexible agents only in the religious committed individuals can destroy the absorbing phase where only $Y$ individuals survive in the long run. On the other hand, in the presence of zealots only in the not religious affiliated subpopulation we observe the persistence of such absorbing state, but the phase where the $Y$ population disappears is not present anymore. Finally, in the presence of inflexible agents in both religious committed individuals and not religious affiliated ones the only collective macroscopic state is the coexistence of all the subpopulations, i.e., we have no phase transitions anymore.


\section*{Acknowledgments}

The author acknowledges financial support from the Brazilian scientific funding agencies Conselho Nacional de Desenvolvimento Cient\'ifico e Tecnol\'ogico (CNPq, Grant 308643/2023-2) and Funda\c{c}\~ao de Amparo \`a Pesquisa do Estado do Rio de Janeiro (FAPERJ, Grant 203.217/2017).


\appendix
\section{Analytical considerations}

In this appendix we will detail some of the analytical calculations considering the mean-field formulation of the model (fully-connected population, section $3.1$).

Let us start considering the long-time limit, $t\to\infty$, in Eq. \eqref{eq1}. Taking $dx/dt = 0$, we obtain two solutions, namely
\begin{eqnarray} \label{app_eq1}
x & = & 0 \\ \label{app_eq2}
z & = & \frac{p-w\,y}{q}
\end{eqnarray}
\noindent
For Eq. \eqref{eq2}, taking $dy/dt = 0$ we also obtain two solutions,
\begin{eqnarray} \label{app_eq3}
y & = & 0 \\ \label{app_eq4}
x & = & \frac{u\,z}{w}
\end{eqnarray}
\noindent
For Eq. \eqref{eq3}, we obtain the following relation,
\begin{equation} \label{app_eq5}
(p-q\,z+u\,z)\,x = u\,z\,(1-z)
\end{equation}

If Eq. \eqref{app_eq4} is valid, Eq. \eqref{app_eq5} gives us two other solutions,
\begin{eqnarray} \label{app_eq6}
z & = & 0 \\ \label{app_eq7}
z & = & \frac{w-p}{u+w-q}
\end{eqnarray}

 Observe that, for the region of parameters where Eq. \eqref{app_eq6} is valid ($z=0$), Eq. \eqref{app_eq4} gives us $x=0$, leading to $y=1$ from the normalization condition. This solution $(x,y,z) = (0,1,0)$ represents the absorbing state of the model, where there are only not religious affiliated individuals in the population for long times, as discussed in section $3.1$.

We considered $z=0$ as a valid solution to obtain the absorbing state. If now we consider Eq. \eqref{app_eq7} valid, Eq. \eqref{app_eq4} gives us another solution for $x$, and after we can consider the the normalization condition to obtain $y$. These solutions are given by
\begin{eqnarray} \label{app_eq8}
x & = & \frac{u\,(w-p)}{w\,(u+w-q)} \\ \label{app_eq9}
y & = & 1 - \frac{(u+w)\,(w-p)}{w\,(u+w-q)}
\end{eqnarray}  

Now we can consider the solution given by Eq. \eqref{app_eq3}. Using this result in Eq. \eqref{app_eq2}, we obtain $z=p/q$. Since this solution is valid for $y=0$, the normalization condition gives us $x=1-p/q$. These results defines another solution for the model, namely $(x,y,z)=(1-p/q,0,p/q)$.

From the previous three solutions, and considering $y$ as the order parameter of the model, we see that we have the solutions $y=0$, $y=1$ and $y$ given by Eq. \eqref{app_eq9}. From Eq. \eqref{app_eq9} we can obtain the two critical points of the model. This solution can be rewriten as $y\sim (p-p_{c}^{(1)})^{\beta}$, that approaches zero for 
\begin{equation}
p_{c}^{(1)} = \frac{w\,q}{u+w}
\end{equation}
\noindent
and present a critical exponent $\beta=1$, as in the  directed percolation problem \cite{dickman}.

The other critical point can be obtained looking again for Eq. \eqref{app_eq9}. This solution achieves the absorbing state where $y=1$ for $p=w$, defining the second critical point,
\begin{equation}
p_{c}^{(2)} = w 
\end{equation}

The above equations define the three distinct solutions discussed in section $3.1$, as well the two critical points of the model.

The local stability of the above equilibrium points can be inferred from the eigenvalues of the Jacobian matrix $J$ obtained from the system of differential equations \eqref{eq1} - \eqref{eq3}. An equilibrium point is locally asymptotically stable if all eigenvalues of $J$ have negative real parts \cite{holmes}. The eigenvalues can be obtained from $det(J-\lambda\,I)=0$, where $I$ is the identity matrix. The Jacobian matrix is given by taking the partial derivatives of Eqs. \eqref{eq1} - \eqref{eq3}, namely

\begin{equation*}
J = 
\begin{bmatrix}
\frac{\partial{\dot{x}}}{\partial{x}} & \frac{\partial{\dot{x}}}{\partial{y}} & \frac{\partial{\dot{x}}}{\partial{z}} \\
\frac{\partial{\dot{y}}}{\partial{x}} & \frac{\partial{\dot{y}}}{\partial{y}} & \frac{\partial{\dot{y}}}{\partial{z}} \\
\frac{\partial{\dot{z}}}{\partial{x}} & \frac{\partial{\dot{z}}}{\partial{y}} & \frac{\partial{\dot{z}}}{\partial{z}}
\end{bmatrix}
\end{equation*}

Taking the derivatives, the matrix becomes

\begin{equation*}
J = 
\begin{bmatrix}
-p + w\,y + q\,z & w\,x & q\,x \\
-w\,y & -w\,x+u\,z & u\,y \\
p - q\,z & -u\,z & -u\,y-q\,x
\end{bmatrix}
\end{equation*}

For the absorbing state solution $(x,y,z) = (0,1,0)$, the eigenvelues of the Jacobian matrix are $\lambda_{1}=0, \lambda_{2}=-u$ and $\lambda_{3}=w-p$. We have $\lambda_{2}<0$ since $u>0$, and $\lambda_{3}<0$ if we have $p>w$. This is exactly the condition we obtained above, i.e., the absorbing state solution is valid for $p>p_{c}^{(2)}=w$.

Another equilibrium is given by $(x,y,z)=(1-p/q,0,p/q)$. The eigenvelues of the Jacobian matrix are $\lambda_{1}=0, \lambda_{2}=p-q$ and $\lambda_{3}=(u+w)(p/q) - w$. We have $\lambda_{2}<0$ for $p<q$ which leads to $x=1-p/q<1$ and $z=p/q<1$, which is expected since $x$ and $z$ are subpopulation densities. We have $\lambda_{3}<0$ for $p<wq/(u+w)$. This is the condition we obtained above, i.e., the solution $(x,y,z)=(1-p/q,0,p/q)$ is valid for $p<p_{c}^{(1)}=wq/(u+w)$. The third solution (coexistence of the three subpopulations $x, y$ and $z$) is then stable for $p_{c}^{(1)} < p < p_{c}^{(2)}$.


\bibliographystyle{elsarticle-num-names}

\end{document}